\newif\ifdraft
\draftfalse

\newif\ifresubmission
\resubmissionfalse

\documentclass[
    aip,
    cha,
    reprint,
    secnumarabic,
    nofootinbib, 
    tightenlines,
    nobibnotes, 
    showkeys, 
    showpacs,
    superscriptaddress,
]{revtex4-1}

\newcommand{\memory}{\ensuremath{Me}}

\newcommand{\zeit}{\ensuremath{t}}

\ifdraft    
\newcommand{\crossout}[1]{{\color{red}\sout{#1}}}
\newcommand{\comment}[1]{{\color{green}[NBB: #1]}}
\else   
\newcommand{\crossout}[1]{}
\newcommand{\comment}[1]{}
\fi 

\ifresubmission

\else

\fi

\newcommand{\figref}[1]{Fig.~\ref{fig/#1}}

\usepackage{natbib}
\usepackage{amsmath,amsfonts,amssymb,amsbsy,amscd,amsgen}
\usepackage{dcolumn}
\usepackage{bm}
\usepackage[pdftex]{graphicx}
\usepackage[caption=false,justification=raggedright,singlelinecheck=false,labelformat=empty]{subfig}
\usepackage{ifthen}

\usepackage{fancyhdr}
\usepackage{siunitx}

\usepackage{xcolor}
\graphicspath{{arXiv-v3/figs/}}
\usepackage[pdftex,colorlinks]{hyperref} 
\hypersetup{
    colorlinks,
    linkcolor={red!50!black},
    citecolor={blue!50!black},
    urlcolor={blue!80!black}
}

\usepackage[T1]{fontenc}
\usepackage[utf8]{inputenc}
\usepackage{color}
\usepackage[normalem]{ulem} 
\usepackage[percent]{overpic}

\begin{document}
\title[]{Crises and chaotic scattering in hydrodynamic pilot-wave experiments}

\author{George Choueiri}%
\thanks{These two authors contributed equally.}
\affiliation{
    Institute of Science and Technology Austria,
    3400 Klosterneuburg, Austria
    }    
\affiliation{
    MIME Department, University of Toledo, Ohio 43606, USA
}

\author{Balachandra Suri}
\thanks{These two authors contributed equally.}
\affiliation{
    Institute of Science and Technology Austria,
    3400 Klosterneuburg, Austria
    }
\affiliation{
    Dept. of Mechanical Engineering, Indian Institute of Science, 
    Bengaluru-560012, India 
}

\author{Jack Merrin}
\affiliation{
    Institute of Science and Technology Austria,
    3400 Klosterneuburg, Austria
    }

\author{Maksym Serbyn}
\affiliation{
    Institute of Science and Technology Austria,
    3400 Klosterneuburg, Austria
    }
    
\author{Björn Hof}
\affiliation{
    Institute of Science and Technology Austria,
    3400 Klosterneuburg, Austria
    }
    
\author{Nazmi Burak Budanur}
\email{nbudanur@pks.mpg.de}
\affiliation{
    Institute of Science and Technology Austria,
    3400 Klosterneuburg, Austria
    }
\affiliation{
    Max Planck Institute for the Physics of Complex Systems,
    01187 Dresden, Germany
}   

\date{\today}

\begin{abstract}
    Theoretical foundations of chaos have have been predominantly laid out for
    finite-dimensional dynamical systems, such as the three-body problem in
    classical mechanics and the Lorenz model in dissipative systems. In
    contrast, many real-world chaotic phenomena, e.g. weather, arise in systems
    with many  (formally infinite) degrees of freedom,  which limits direct
    quantitative analysis of such systems using chaos theory. In the present
    work, we demonstrate that the hydrodynamic pilot-wave systems offer a bridge
    between low- and high-dimensional chaotic phenomena by allowing for a
    systematic study of how the former connects to the latter. Specifically, we
    present experimental results which show the formation of low-dimensional
    chaotic attractors upon destabilization of regular dynamics and a final
    transition to high-dimensional chaos via the merging of distinct chaotic
    regions through a crisis bifurcation. Moreover, we show that the post-crisis
    dynamics of the system can be rationalized as consecutive scatterings from
    the nonattracting chaotic sets with lifetimes following exponential
    distributions.  
\end{abstract}

\maketitle

\begin{quotation}
        Hydrodynamic pilot-wave systems formed by a milimetric droplet
        bouncing on the vertically vibrating bath of the same fluid are
        primarily studied for their analogies to quantum phenomena. These
        intriguing analogies are rooted in the rich dynamics that arise from the
        interaction of the droplet with the surface waves that it generates at
        each impact. Here, we experimentally study one such system formed by a single droplet
        bouncing on a circularly shaped vibrating bath and
        uncover the qualitative changes in the system's dynamics as a control
        parameter is slowly varied. In particular, we present compelling
        evidence that some of these changes correspond to so-called ``crisis
        bifurcations'' wherein a chaotic attractor loses stability giving way to
        a discontinuous change in the dynamics upon a small variation of the
        control parameter. Finally, we show that the complex dynamics of 
        the system that follows a merger of previously distinct chaotic sets can 
        be understood as scatterings from chaotic repellers with distinct 
        physical properties. 
\end{quotation}

\section{Introduction}
\label{s-intro}

When a fluid bath vibrates vertically with an amplitude above a critical value,
the surface undergoes an instability leading to the spontaneous formation of the
so-called Faraday waves.\cite{faraday1837on,benjamin1954stability} Nearly two
decades ago, Yves Couder and coworkers\cite{couder2005from,couder2005walking}
showed that when a bath of silicone oil is vertically vibrating with an
amplitude slightly below this point of instability, a millimeter-sized droplet
of the same fluid can bounce vertically and ``walk'' (move horizontally) on the 
surface without coalescing due to the presence of a thin
air layer between the droplet and the bath surface. In this regime, the impulse
that the droplet experiences at each impact is determined by the surface
topography, which itself is shaped by the waves generated at previous bounces.
Although these waves do not persist and decay
exponentially, they do so at a rate that vanishes as the vibration amplitude
approaches the Faraday threshold, thus retaining a
``memory''\cite{eddi2011information} of the droplet's trajectory.
Soon after the initial demonstrations of bouncing and
walking, Couder and coworkers realized macroscopic
analogs to quantum mechanical phenomena such as tunneling 
and orbital quantization 
on these setups.\cite{eddi2009unpredictable,fort2010pathmemory} 
Their results were extended in subsequent studies\cite{cristeaplaton2018walking,harris2013wavelike,saenz2020hydrodynamic,
durey2020speed,saenz2020hydrodynamic,tadrist2020predictability}
and the setups were named 
``hydrodynamic pilot-wave systems''\cite{bush2015pilotwave} due to their 
reminiscence of the de Broglie-Bohm interpretation of quantum mechanics. 
Today, the exploration of analogies between pilot-wave hydrodynamics and quantum 
phenomena continues to be an active area of research; see the recent 
review by Bush and Oza.\cite{bush2020hydrodynamic}

Over the past two decades, pilot-wave hydrodynamics have also 
received considerable attention for their dynamical properties.\cite{rahman2020walking}
Hydrodynamic pilot-wave systems can
be viewed from two different theoretical perspectives that are compatible with
one another. If one approximates the droplet as a point particle and takes its position and momentum as state variables, then the corresponding dynamical system is non-Markovian since the time-evolution of the system not only depends
on its present state but also on its history which determines the surface
topography.\cite{eddi2011information,oza2013trajectory} Alternatively, if one
takes the droplet and fluid bath as a whole and includes the bath's
configuration in the description of the system's
state\cite{perrard2018transition,budanur2019state} then the knowledge of the
present state becomes sufficient for predicting its future, thus rendering the
system Markovian. In this latter approach, the hydrodynamic pilot-wave setup
constitutes an infinite-dimensional dynamical system since the surface field
approximated as a continuum introduces infinitely-many degrees of freedom. In
the following, we adopt this infinite-dimensional Markovian view in order to
explain how complex dynamics can arise in a hydrodynamic pilot-wave system. 

The view of complex hydrodynamic phenomena as those arising in
infinite-dimensional dynamical systems can be traced to Hopf's early
mathematical work in turbulence.\cite{hopf1948mathematical} Over the past thirty
years, this approach enjoyed a resurgent interest mainly due to the advancements
in computing hardware and numerical algorithms that made searching for simple
time-invariant solutions such as equilibria and periodic orbits in
fully-resolved Navier--Stokes simulations
feasible.\cite{kawahara2012significance,graham2021exact} Besides numerical
simulations, the influence of invariant solutions on weakly turbulent
flows was also demonstrated in
experiments.\cite{hof2004experimental,suri2020capturing} One of the
frequently-stated goals of this research field is building low-dimensional
models of turbulence using (unstable) periodic orbits as building
blocks.\cite{cvitanovic2013recurrent} Although several
papers\cite{cvitanovic2010geometry,chandler2013invariant,budanur2017relative,
suri2020capturing,krygier2021exact} illustrated the resemblance of turbulent
flows and periodic orbits in different fluid flows, building quantitatively
accurate models based on periodic orbits has only been possible in
highly-restricted configurations.\cite{yalniz2021coarse} In this paper, we
explore an alternative approach to this problem, namely one that takes the
chaotic repellers as building blocks as opposed to periodic orbits, and
demonstrate its success through our analysis of hydrodynamic pilot-wave
experiments. 

Finite-lifetime chaotic motion can be observed in a variety of settings,
such as chemical reactions,\cite{noid1986fractal}  
advection of suspended particles,\cite{sommerer1996experimental} and dynamics of
transitionally  turbulent flows.\cite{hof2006finite} Among these, a well-studied
phenomenon is chaotic scattering, that is the scattering of a particle from the
neighborhood of a nonattracting chaotic set in the phase space of a 
system.\cite{eckhardt1987fractal,eckhardt1988irregular,gaspard1989scattering,ott2002chaos}
Although initially studied in open billiard 
systems,\cite{eckhardt1987fractal,eckhardt1988irregular,gaspard1989scattering} 
chaotic scattering found applications in various other systems with chaotic 
repellers, for recent examples see 
Refs.\cite{goodman2007chaotic,goodman2008chaotic,goodman2015mechanical}. In the 
present paper, we demonstrate that the chaotic dynamics of a hydrodynamic 
pilot-wave system can be decomposed into chaotic repellers formed around 
distinct periodic solutions of the system. Consequently, we show that the 
observed dynamics can be understood as consecutive scatterings from these 
repellers.

Our experiments consist of a single droplet bouncing on a bath with variable
topography that introduces a radially confining force, which is known to enable
chaotic dynamics in hydrodynamic pilot-wave
systems.\cite{perrard2014chaos,tambasco2016onset,perrard2018transition} In this
setup, we slowly change the control parameter to uncover the series of
bifurcations that lead to the formation of the system's chaotic attractor. In
particular, we observe crisis bifurcations,\cite{grebogi1982chaotic} i.e., discontinuous changes of the
system's attractor upon small changes of the control parameter. In the final
chaotic regime, we study lifetime statistics of dynamics in different parts of
the attractor and show that they have exponential tails as expected for chaotic
repellers.\cite{lai2011transient} Our results suggest that the final crisis
bifurcation that leads to the formation of the system's attractor is mediated by
a chaotic repeller, in contrast to well-studied \cite{
grebogi1982chaotic,grebogi1987critical,ott2002chaos} crisis phenomena that take
place when a chaotic attractor of a continuous-(discrete-)time system collides
with a periodic orbit (fixed point) of saddle type. We, thus, argue that our
findings open new theoretical questions for high-dimensional chaotic systems. 

This article is structured as follows. In section \ref{sec:methods}, we
discuss the experimental setup, a symmetry-reducing transformation of
experimental data, and the construction of Poincar\'e sections. 
In section \ref{sec:results} we examine how  droplet
dynamics change as the memory $Me$ of the vibrating fluid bath is varied. We also
rationalize our findings using tools from chaos theory. Finally, we discuss the
significance of our findings in the broader context of hydrodynamic quantum
analogs as well as fluid turbulence in \ref{sec:remarks}.
    
\section{Methods}
\label{sec:methods}
\subsection{Experimental setup}

Our setup consists of a computer-controlled electromagnetic shaker
on which a bath containing silicone oil is mounted and a camera above records the
dynamics of the bouncing droplet. 
As shown in \figref{setup}, the bath is in the shape of a circular corral formed  
by concentric cylinders with a deep inner
section surrounded by a shallow damping (overflow) region; a configuration which
was shown to yield an effective radially-confining 
force.\cite{harris2013wavelike,cristeaplaton2018walking,durey2020faraday}
All experimental runs are performed by setting the bath's vibration frequency
to $f_0 = 75\,\si{Hz}$ and  using a single droplet with a diameter 
$D = 0.85 \pm 0.05\,\si{mm}$. The bath acceleration $\gamma$ is varied to adjust the \textit{memory}
\begin{equation}
	\memory = (1 - \gamma / \gamma_F)^{-1} \, . \label{e-Memory}
\end{equation}
{$\memory$}  is a dimensionless control parameter that is proportional to the
damping time of the surface waves \cite{eddi2011information} and tends to
infinity as the bath acceleration $\gamma$ approaches the Faraday instability
threshold at $\gamma_F$. 
    In our experiments, we adjust $\gamma$ according to \eqref{e-Memory} such that 
    \memory\ is varied in approximately equal steps and study the changes in the 
    system's behavior.
In presenting our results, we 
    convert length and time to dimensionless quantities by measuring them in 
    units of the  Faraday wavelength 
$\lambda_F = 5.3\,\si{mm}$  and time $t_F=2/f_0$, respectively.
The droplet trajectories are reconstructed via image processing for 
detecting the droplet center in each frame recorded by the camera (\figref{setup}{B}), 
and the droplet's instantaneous velocity is estimated by computing the time-derivative 
of the cubic splines that fit these trajectories.
Further experimental details can be found in the Appendix. 

\begin{figure}
	\centering
	\begin{overpic}[width=3.2in]{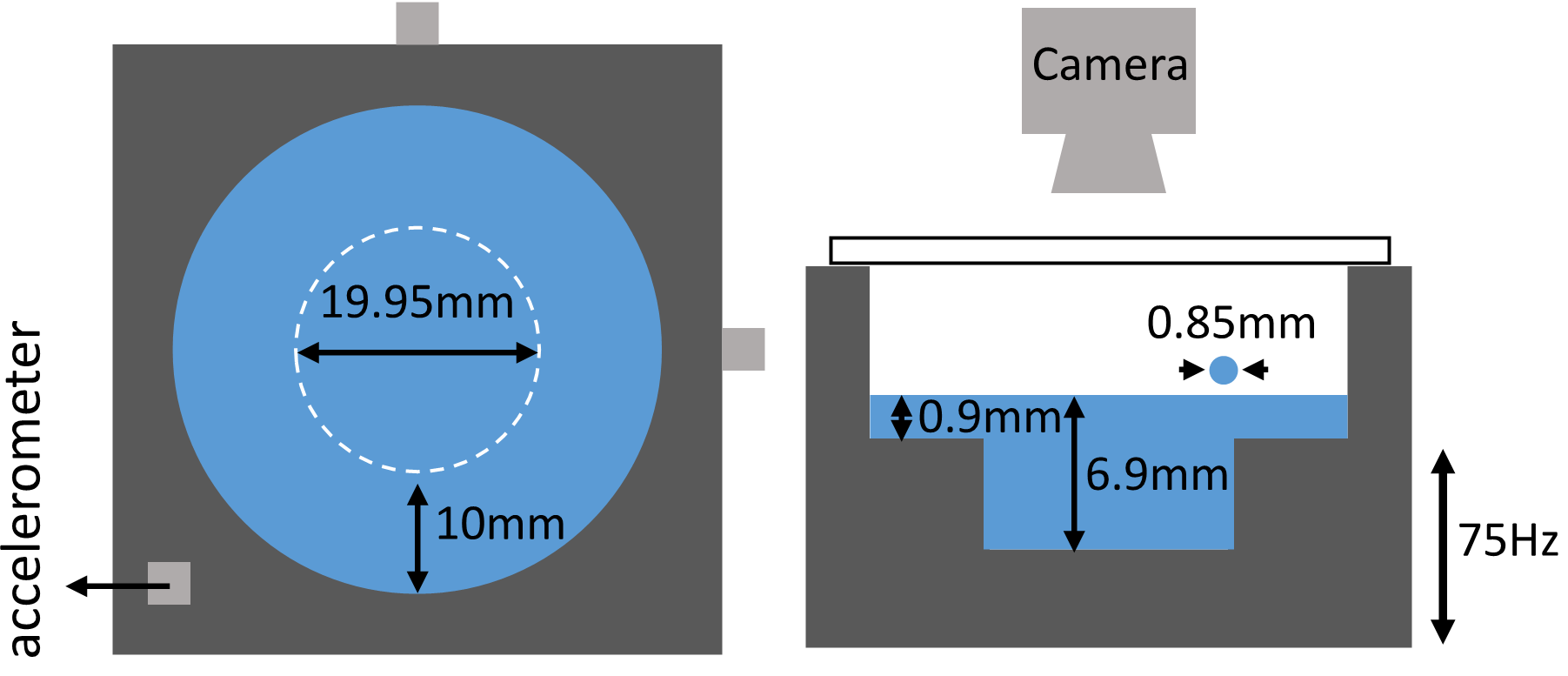}
		\put (2,40) {A.}
		\put (49,40) {B.}
	\end{overpic}
	\caption{
        \label{fig/setup} Schematic of the experimental setup. (A) Top view
	    showing the corral (white dashed circle), the overflow region, and
	    accelerometers (gray squares). (B) Side view illustrating fluid layer and
    	imaging configurations. The center of the corral is chosen as the
	    origin of the coordinate system.
        }
\end{figure}

\subsection{Symmetry reduction}

\begin{figure*}
	\centering
\begin{overpic}[width=0.235\textwidth]{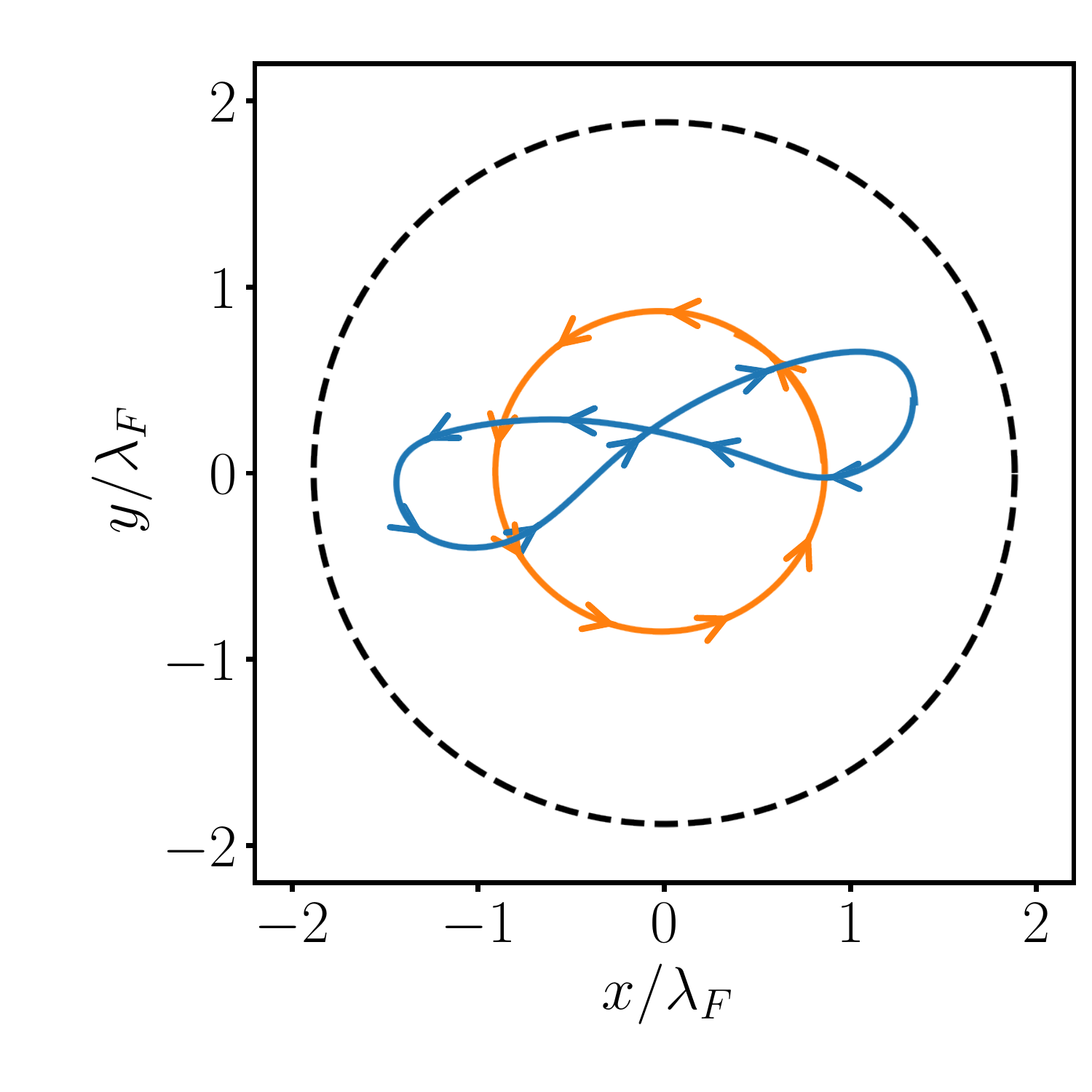}
	\put (15,100) {A. \qquad $\memory = 20$}
	\put(-5,40){\rotatebox{90}{Lab-frame}}
\end{overpic}\,
\begin{overpic}[width=0.235\textwidth]{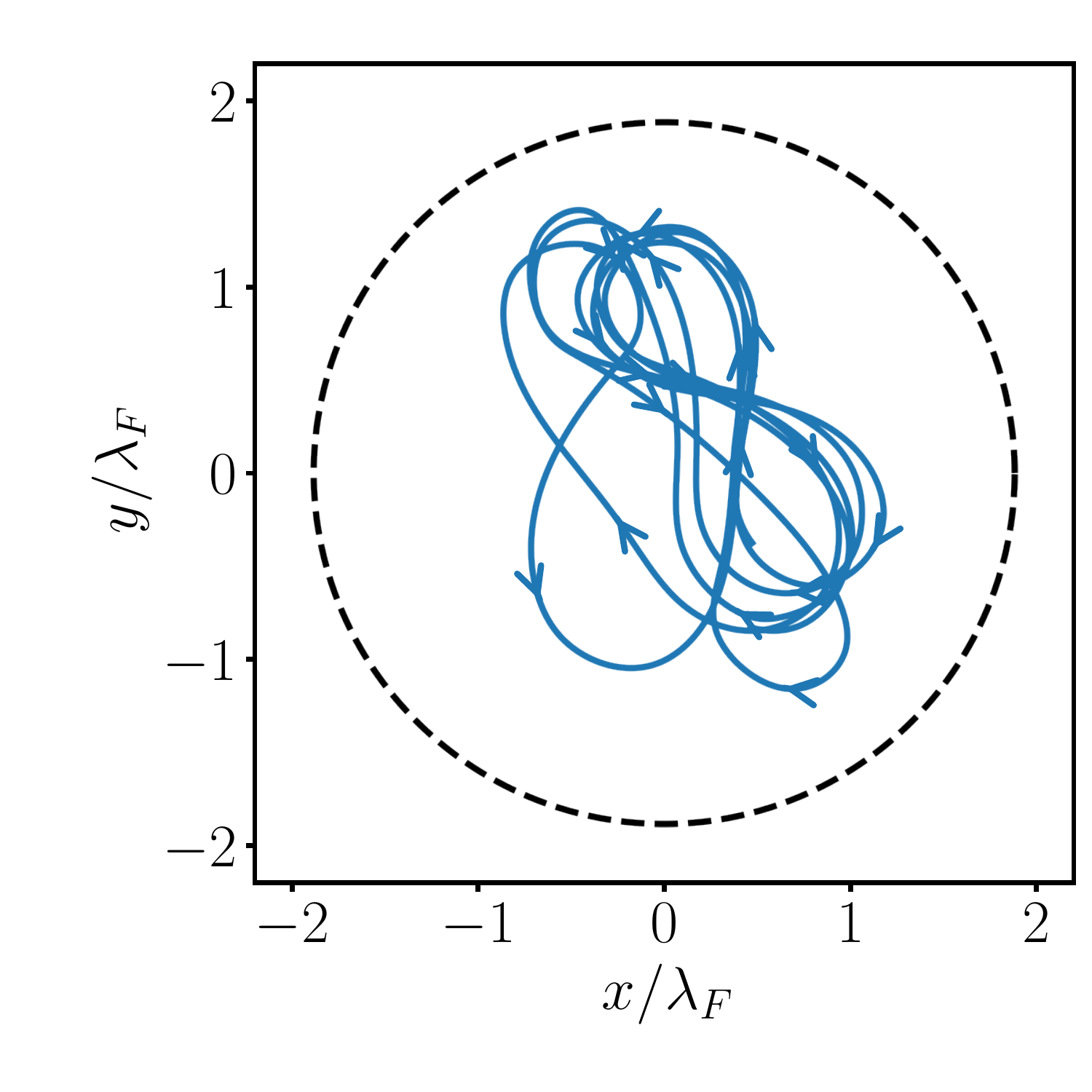}
	\put (15,100) {B. \qquad $\memory = 26.48$}
\end{overpic}\,
\begin{overpic}[width=0.235\textwidth]{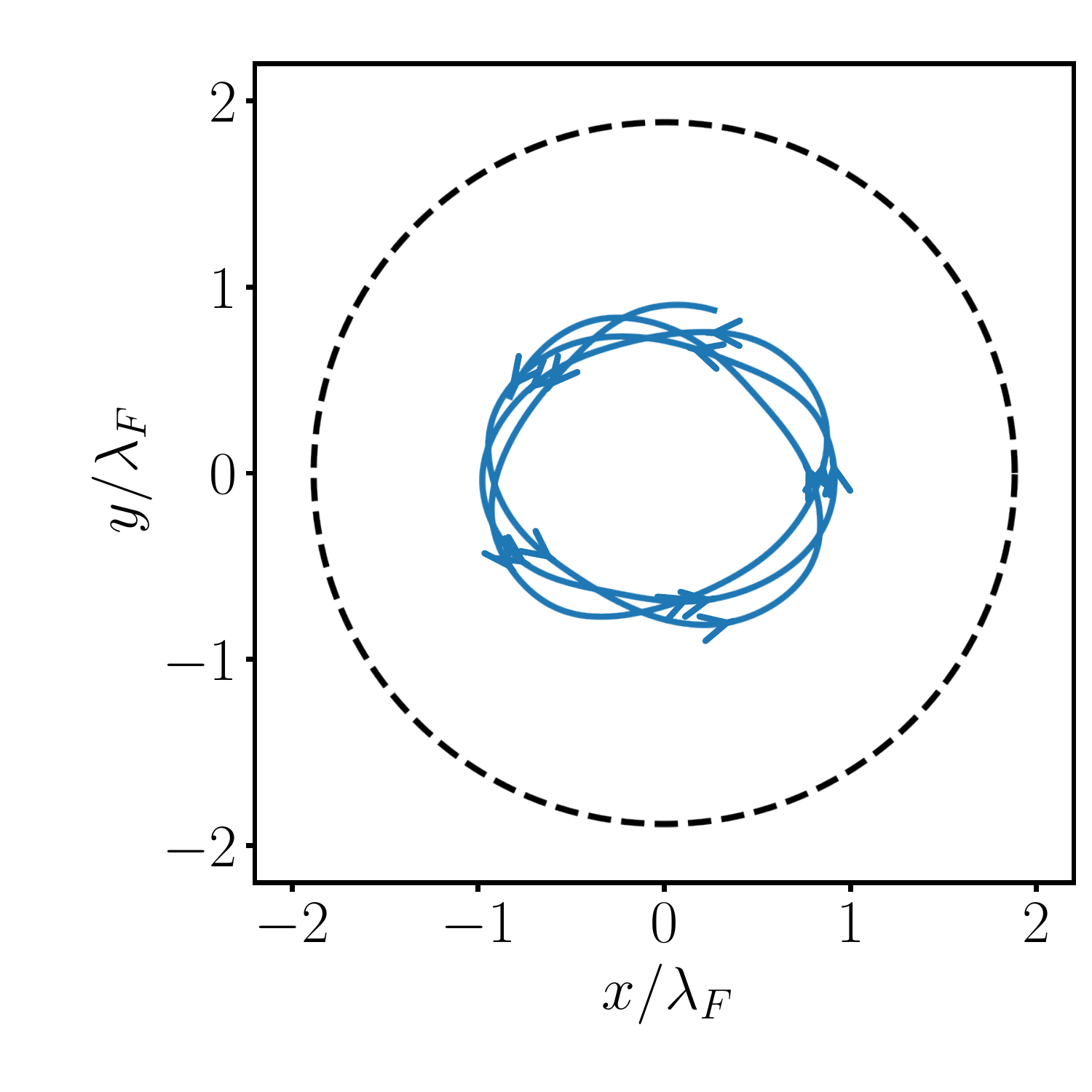}
	\put (15,100) {C. \qquad $\memory = 31.48$}
\end{overpic}\,
\begin{overpic}[width=0.235\textwidth]{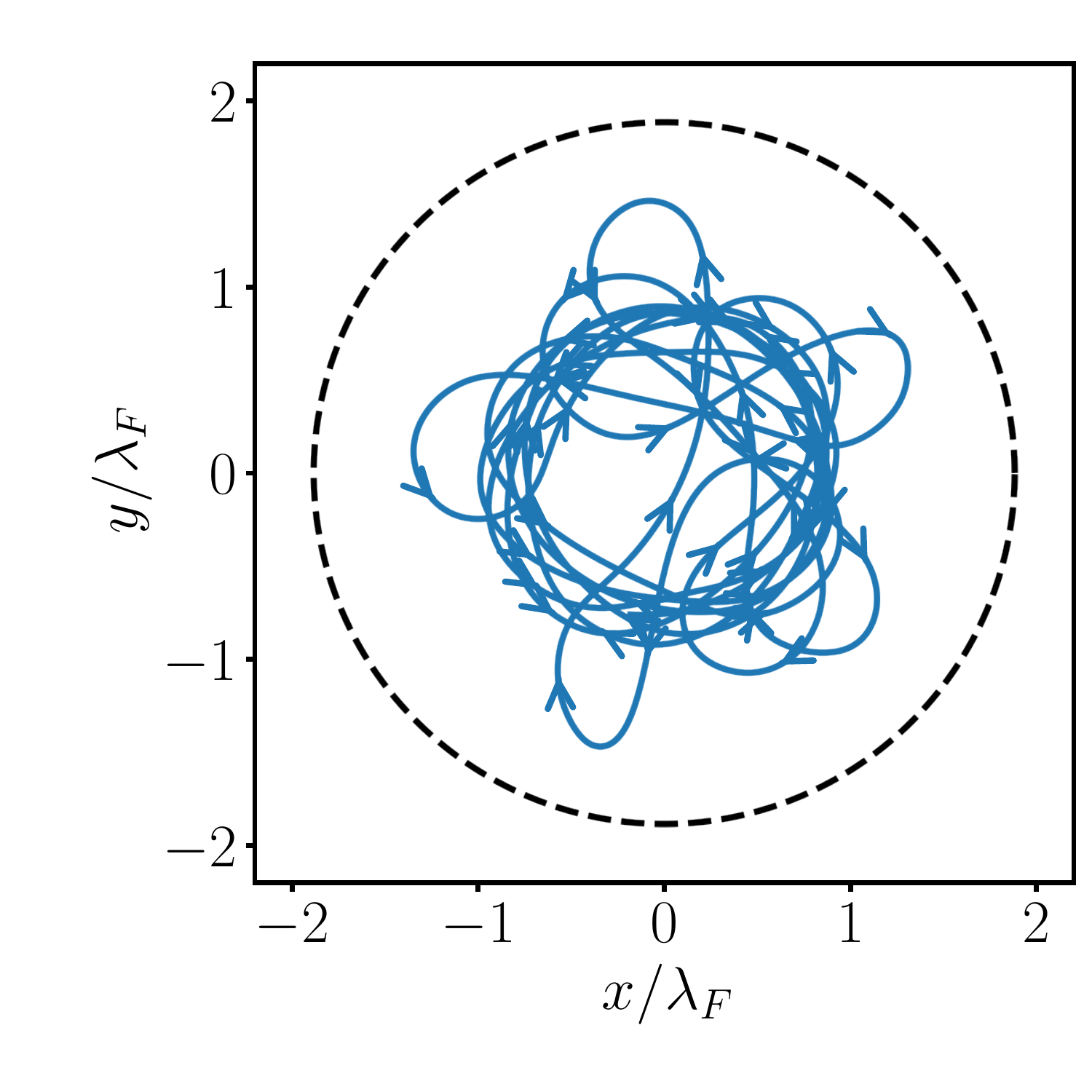}
	\put (15,100) {D. \qquad $\memory = 31.86$}
\end{overpic} \\ \vspace{0.01\textwidth}
\begin{overpic}[width=0.235\textwidth]{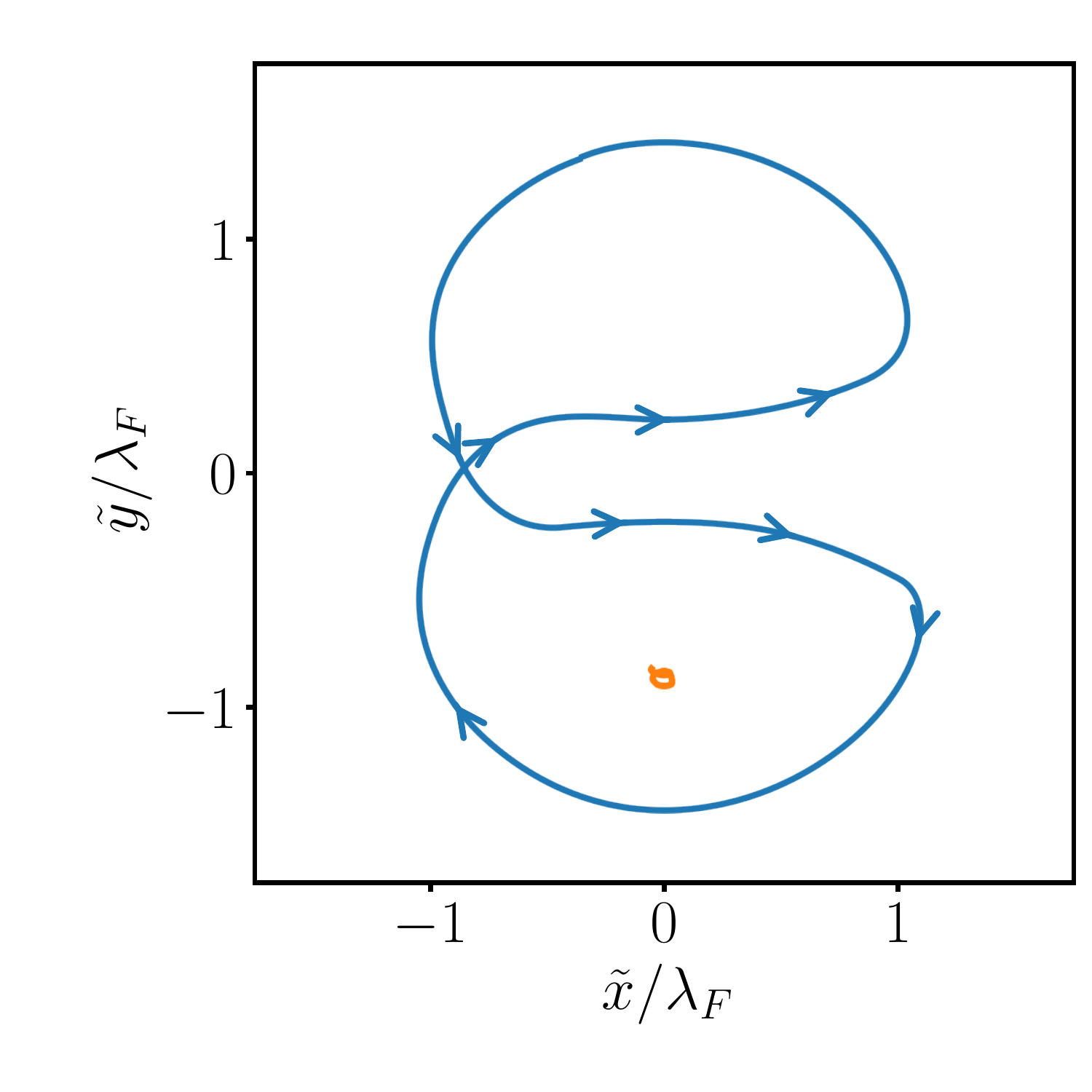}
	  \put(-5,25){\rotatebox{90}{Symmetry-reduced}}
\end{overpic}\,
\begin{overpic}[width=0.235\textwidth]{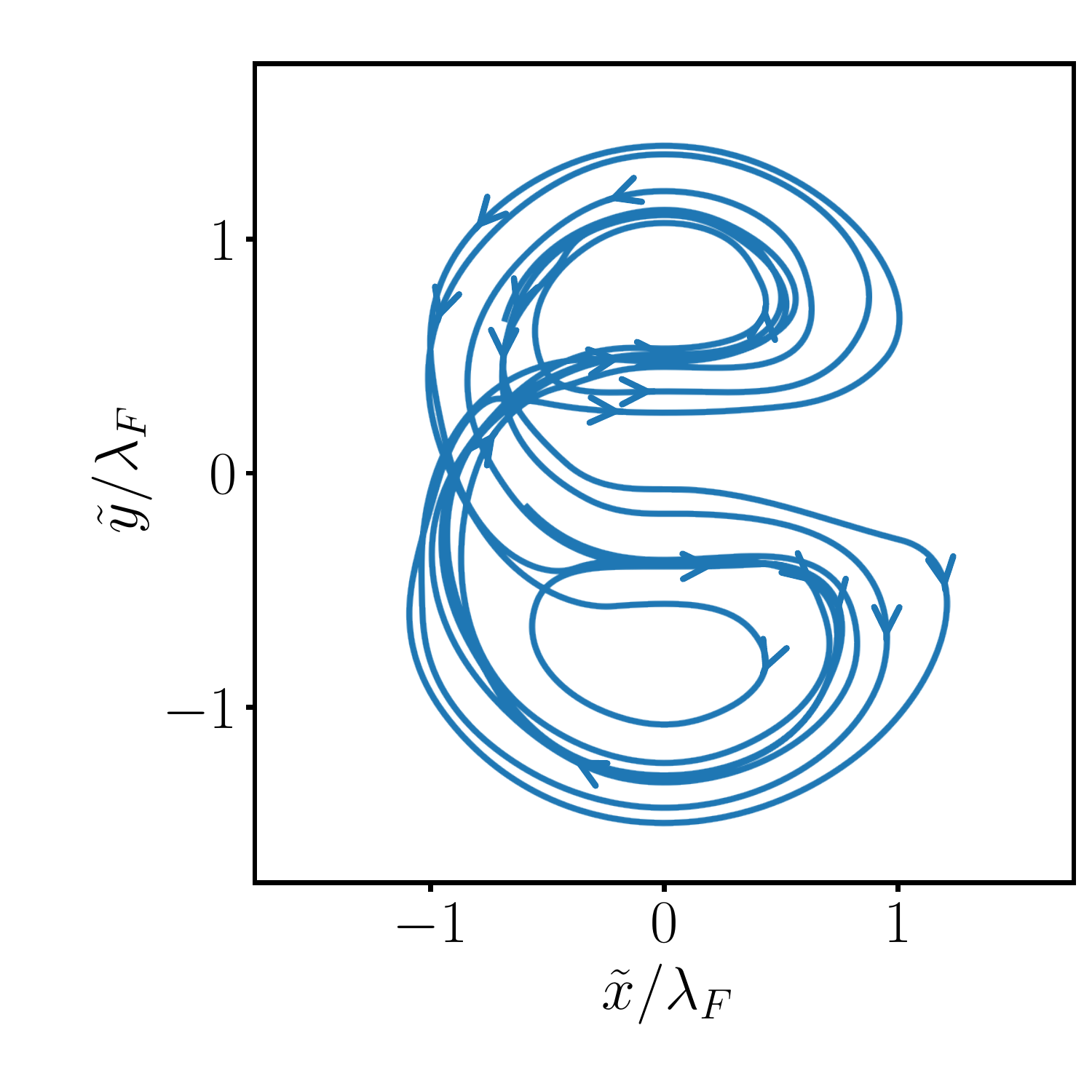}
\end{overpic}\,
\begin{overpic}[width=0.235\textwidth]{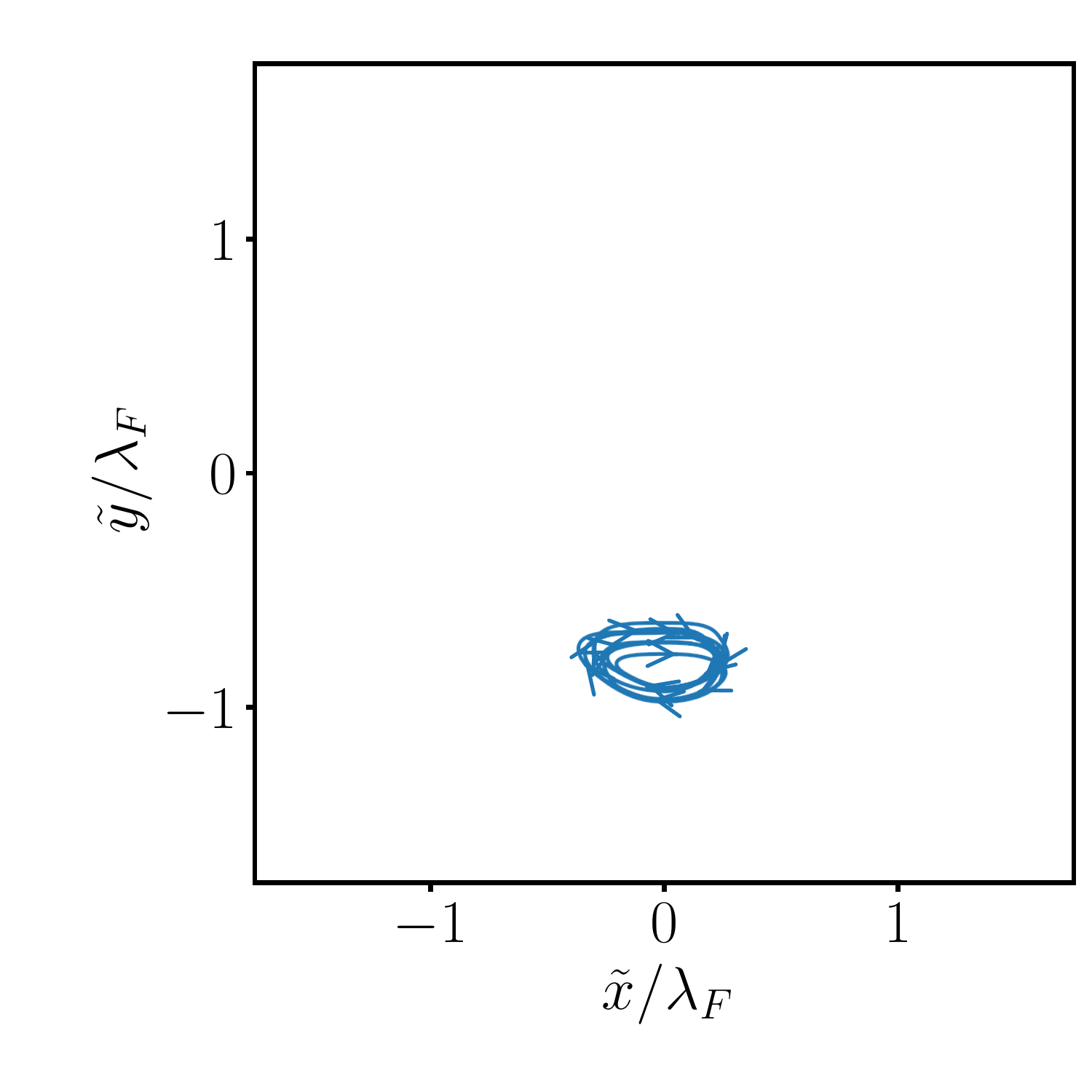}
\end{overpic}\,
\begin{overpic}[width=0.235\textwidth]{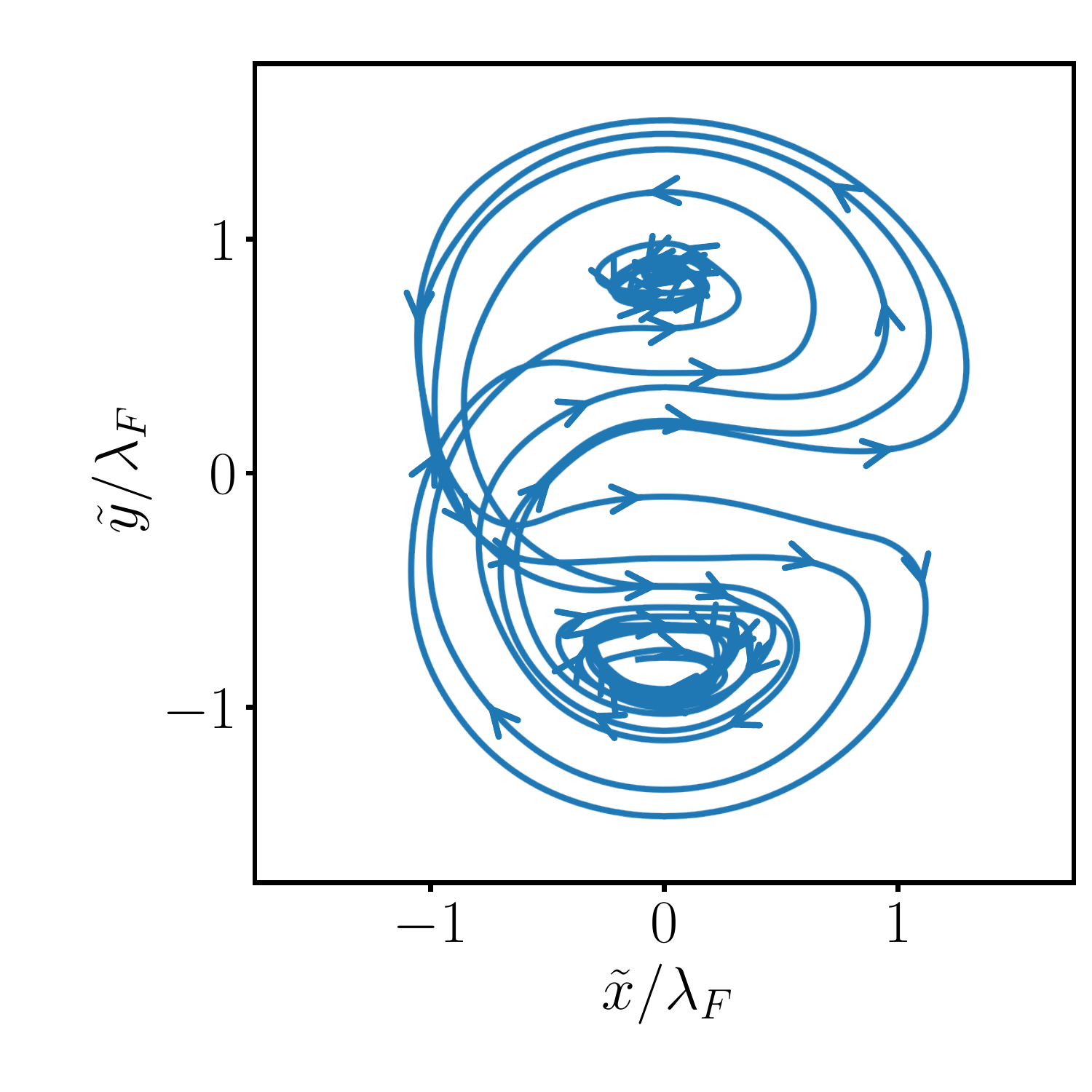}
\end{overpic}
	\caption{Experimentally measured droplet trajectories
			 where the top row shows the trace of the drop
			 in the lab and the bottom row shows their
			 symmetry-reduced representations.
			 The arrow heads indicate the direction of time
			 and the dashed circles show the boundary of
			 the corral.
			 (A) Circle (orange) and lemniscate (blue)
			 at $\memory = 20$,
			 (B) lemniscate chaos at $\memory = 26.48$,
		     (C) oval chaos at $\memory = 31.48$,
		     (D) chaos $\memory = 31.86$.
		\label{fig/trajectories}
	}
\end{figure*}

\figref{trajectories} (top) shows examples of reconstructed droplet trajectories 
in our experiments at different \memory.
While visualizations in the laboratory frame are illustrative, 
they contain redundant information due to symmetries of the corral, i.e., azimuthal 
reflection and rotations.
As a result of the reflection symmetry, for each
clockwise rotating orbit, the system also
exhibits a dynamically equivalent  counter-clockwise rotating one. In the case
of the circular orbit in \figref{trajectories}{A} for instance, the clockwise
and counter-clockwise orbits can be distinguished using the sign of their
angular momentum $L = m ({\bf x} \times {\bf v})$, where $m$ is the mass of the
silicone oil droplet, and
${\bf x}(t) = [x(t),y(t)]$,  ${\bf v}(t) = [v_x(t),v_y(t)]$ are its
instantaneous two-dimensional position and velocity, respectively.
These two orbits trace approximately the same trajectory on the $xy$-plane.
The continuous rotation symmetry poses a different challenge; any non-circular
trajectory (e.g., lemniscate in \figref{trajectories}{A}) can be rotated by an 
arbitrary angle about the origin to generate a dynamically equivalent
trajectory at a different orientation. In order to eliminate this degeneracy, 
we apply Budanur and Fleury's \cite{budanur2019state} continuous 
symmetry reduction method to our experimental data. The basic idea of this method 
is to set the polar angle in the velocity plane of the pilot-wave system's 
state space to a fixed value so that each rotation-equivalent trajectory
is mapped to a single representative. Formally, this transformation is applied to 
all dynamical degrees of freedom, including those encoding the state of the bath. 
In the present experimental case, we neglect the bath's degrees of freedom which
are not essential for our discussion to follow and transform the
measured coordinate ${\bf x}(\zeit)$ and velocity
${\bf v}(\zeit)$ as
\begin{equation}
	\tilde{\bf x}(\zeit) = R(-\tilde{\theta}(\zeit)){\bf x}(\zeit) \quad\mbox{and}\quad
	\tilde{\bf v}(\zeit) = R(-\tilde{\theta}(\zeit)){\bf v}(\zeit) \,,
	\label{symm_red}
\end{equation}
where $\tilde{\theta}(\zeit) = \arg (v_x (\zeit) + i v_y (\zeit))$ is the instantaneous 
polar angle in the velocity plane 
and
$R (\theta)$ is the $2\times2$ matrix 
\begin{equation}
    R (\theta) = 
        \begin{bmatrix}
            \cos \theta & - \sin \theta \\ 
            \sin \theta & \cos \theta 
        \end{bmatrix}
\end{equation}
whose action rotates a two-dimensional vector in counterclockwise 
direction by $\theta$.

By fixing the phase of the measurements on the velocity plane to $0$,
the transformation \eqref{symm_red} maps all-rotation equivalent measurements
to one with $\tilde{v}_x = \|{\textbf{v}}\|$ and $\tilde{v}_y = 0$, hence,
performs a symmetry reduction.
Note that the transformation \eqref{symm_red} is defined as long
as the speed
$\|\textbf{v}(\zeit)\|$ of the droplet does not vanish, which is the case for
the dynamical regime of interest, see Fig.~\ref{fig:vpdf} of the appendix. 
We note that the analogous transformation that maps coordinates $\bf{x}$
to $(\| \bf{x} \|, 0)$ cannot be applied since this transformation is 
 singular at the origin
$x=y=0$, which is approached by the lemniscate-shaped trajectories
(\figref{trajectories}{A,B}).

\figref{trajectories}\ (bottom row) shows the symmetry reduced
trajectories corresponding to the panels above. 
As seen in \figref{trajectories}{A} (orange), symmetry reduction
maps the circular trajectory to approximately a point 
corresponding to the one at which the droplet velocity is 
in positive $x$ direction. 

In the case of the lemniscate orbit, the trajectory 
\figref{trajectories}{A} and all of its rotation copies are mapped to the
symmetry-reduced lemniscate shown in the bottom panel.
Another feature of our symmetry reduction can be understood by
noting the apparent symmetry of the symmetry-reduced lemniscate in
\figref{trajectories}{A} (bottom)
with respect to the $\tilde{y}=0$ line. This follows directly from the fact that
the azimuthal reflection symmetry of our pilot-wave system is equivalent to the
transformation $\tilde{y} \rightarrow -\tilde{y}$ after symmetry reduction.
Consequently, the lower and upper halves of the $\tilde{x}\tilde{y}$-plane
correspond to droplet motions with positive and negative angular momenta,
respectively. This fact is also observed in the circular
(\figref{trajectories}{A}, bottom) and oval 
(\figref{trajectories}{C}, bottom) trajectories,
where the counterclockwise-rotating ($L > 0$) trajectories are confined to the
lower half of the $\tilde{x}\tilde{y}$-plane after symmetry reduction.

\subsection{Reduction to a Poincar\'e section}

In order to identify the sequence of bifurcations, we define the Poincar\'e
section \cite{strogatz2018nonlinear} as the position-velocity pairs
$(\tilde{\bf x}_{\mathcal{P}}, \tilde{\bf v}_{\mathcal{P}})$ on the
symmetry-reduced trajectories that satisfy 
the half-plane condition
\begin{equation}
    \tilde{\bf x}_{\mathcal{P}} \cdot [1, 0] = 0 \quad \& \quad 
    \tilde{\bf v}_{\mathcal{P}} \cdot [1, 0] > 0 \,, 
    \label{poinc_sect}
\end{equation}
which corresponds to taking the 
intersections of the trajectories $(\tilde{\bf x}(t), \tilde{\bf v}(t))$ with 
the $\tilde{y}$ axis in the positive-$\tilde{x}$ direction.
We chose this section since it is intersected by all 
of the symmetry-reduced trajectories (\figref{trajectories}, bottom row) that 
we observe. In our results to follow, we approximated these intersections as 
linear interpolations between the symmetry-reduced data points at times 
$t_i$ and $t_{i+1}$ with $\tilde{x}(t_{i}) < 0$ and $\tilde{x}(t_{i+1}) > 0$.

Because we only measure droplet trajectories, our Poincar\'e section 
\eqref{poinc_sect} in the space of symmetry-reduced trajectories 
$(\tilde{x}, \tilde{y}, \tilde{v}_x, 0)$
corresponds to a two-dimensional plane $(0, \tilde{y}, \tilde{v}_x, 0)$. 
Consequently, the Poincar\'e section \eqref{poinc_sect} retains only a 
subset of the system's state due to our neglect of the bath's degrees 
of freedom. Formally, an equivalent Poincar\'e section could have been 
defined for the infinite-dimensional state measurements containing 
the bath's degrees of freedom as done for the numerical data 
by Budanur and Fleury.\cite{budanur2019state} 
The section \eqref{poinc_sect} should be understood as a projection of 
that Poincar\'e section onto a two-dimensional plane, and as we shall 
demonstrate next, the retained information is sufficient for our 
analysis. 

\section{Results {\& Discussion}}\label{sec:results}
In the following, we present the results of a parametric study wherein we vary 
\memory\ in small increments to follow the changes in the 
symmetry-reduced dynamics. 

\subsection{Crisis Bifurcations}

\begin{figure}
	\centering
	\includegraphics[width=\linewidth]{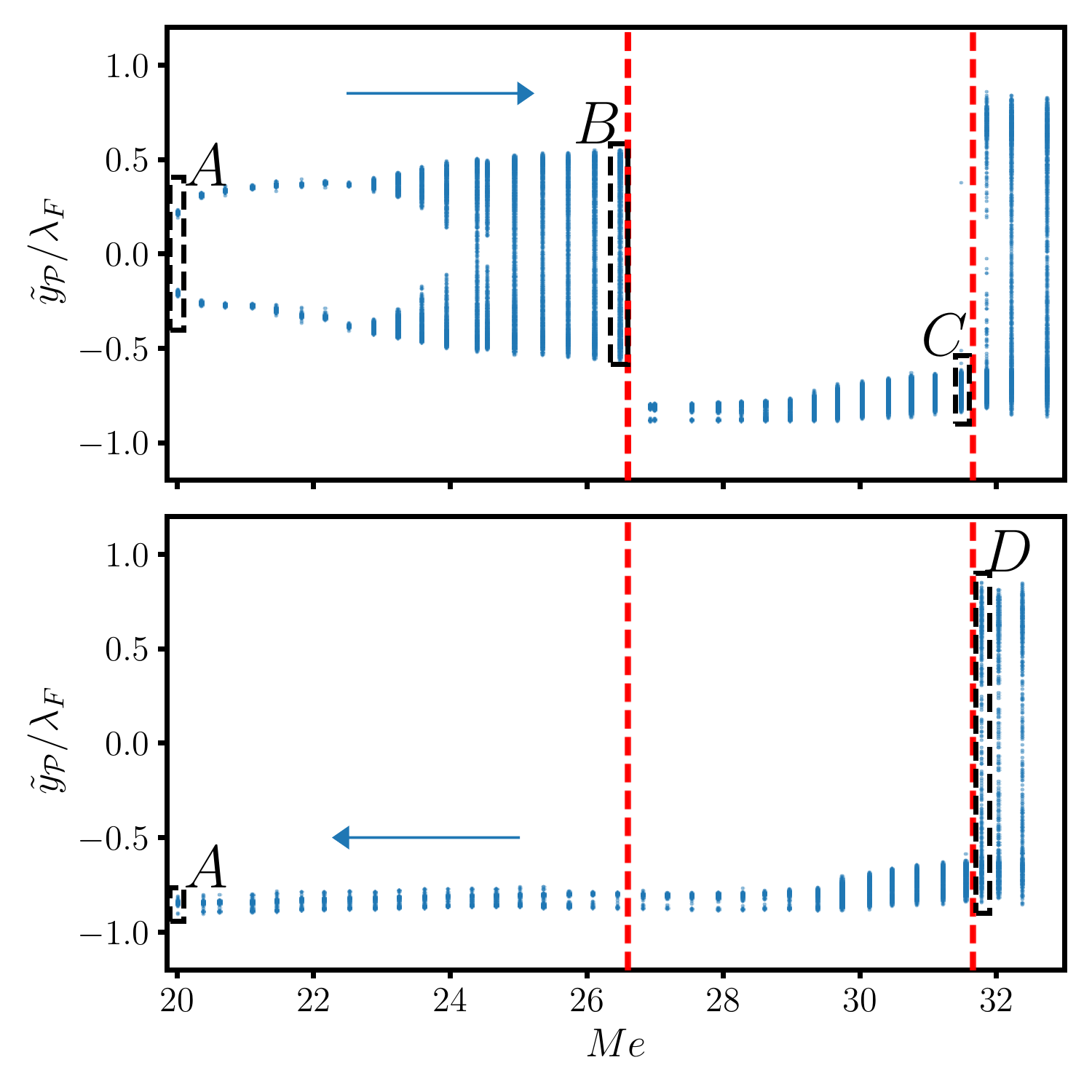}
	\caption{
		Orbit diagram generated by marking the intersections of the
		symmetry-reduced trajectories with the Poincar\'e section \eqref{poinc_sect}.
		Top panel shows the parameter sweep with increasing  
        $\memory$ and bottom panel for the opposite direction as indicated 
        by the arrows. Vertical dashed red lines at 
        $\memory = 26.6$ and  $\memory = 31.65$ mark approximate parameter 
        values at which the crisis  bifurcations take place.
		The labels (A--D) next to the dashed rectangles indicate
		the corresponding trajectories shown in \figref{trajectories}.
	}
	\label{fig/scatterOrbitDiag}
\end{figure}

\figref{scatterOrbitDiag} shows the orbit diagram which we obtained by recording
the symmetry-reduced droplet trajectories' intersections with the Poincar\'e
section \eqref{poinc_sect} starting from the lemniscate trajectory at $\memory =
20$ and increasing $\memory$ in small steps of $\Delta \memory \approx 0.36$ up
to $\memory = 32.8$ and reversing the direction of parameter sweep afterwards.
Top (bottom) panels correspond to the experimental run where \memory\ was
increased  (decreased). 

For the stable lemniscate solution at $Me\approx 20$, we
obtain two (localized) intersections on the Poincar\'e section at
$\tilde{y}_{\mathcal{P}}/\lambda_F\approx \pm 0.025$. Upon increasing the 
\memory, this solution loses stability and a chaotic lemniscate attractor 
(\figref{trajectories}{B})
takes its place. This transition to chaos results in the expansion of the
$\tilde{y}_{\mathcal{P}}$ interval spanned by the intersections in \figref{scatterOrbitDiag} 
(top panel),
starting at $\memory \approx 22.5$. Upon further increase in  \memory, the
chaotic lemniscate attractor loses stability at $\memory \approx 26.6$, after
which we observe circular trajectories similar to the orange curve in
\figref{trajectories}{A}. On the orbit diagram, the corresponding
intersections with the Poincar\'e section are localized at $\tilde{y}_{\mathcal{P}}/\lambda_F\approx
-0.08$. When we increase \memory\ further, we observe modulations to the
circular trajectories and the formation of a new \textit{oval} chaotic attractor
(\figref{trajectories}{C}). This is once again indicated by the gradual
expansion of the markers in the orbit diagram at $Me \approx 29.5$. Finally, we
observe a sudden expansion of the attractor at $\memory \approx 31.65$, which
results in droplet trajectories intermittently switching between oval motions
and lemniscates as shown in \figref{trajectories}{D}. Qualitative features of
this chaotic motion remain unchanged for the higher \memory\ values shown in
\figref{scatterOrbitDiag}. Decreasing \memory\ from this point (bottom panel of
\figref{scatterOrbitDiag}), however, yields a different scenario. Although at
$\memory \approx 31.65$, the dynamics fall back onto the oval chaos
(\figref{trajectories}{C}), we do not recover the lemniscates upon further
decrease of \memory . Rather, we follow the branch of stable circular
trajectories down to $\memory = 20.0$ as shown in \figref{scatterOrbitDiag}. 
In other words, for the parameter interval $\memory \in [20.0, 26.6]$, we have a
multistable system with distinct branches of stable dynamics that can be
observed depending on how the system is initiated.

Sudden changes in the orbit diagram \figref{scatterOrbitDiag} indicated by the
dashed red lines can be understood as crisis bifurcations \cite{grebogi1982chaotic} at
which the attractor of a nonlinear system undergoes a discontinuous change upon
a small variation of the control parameter. The first of these bifurcations takes
place when the lemniscate chaos (\figref{trajectories}{B}) loses
stability at  $\memory \approx 26.6$. This so-called \textit{boundary crisis}
\cite{grebogi1982chaotic} generically takes place when a chaotic set intersects with its
basin boundary. Usually, this boundary is the stable manifold of another
solution, such as a periodic orbit. Naturally, in an experimental study, we 
cannot probe unstable solutions. Nevertheless, we note
that multistability of the system for $\memory \in [20.0, 26.6]$ is in
agreement with this scenario.
The second crisis bifurcation takes place at $\memory = 31.65$, when
the chaotic attractor with oval-shaped trajectories (\figref{trajectories}{C})
suddenly expands into the upper-half of the $\tilde{y}_{\mathcal{P}}$ axis in
\figref{scatterOrbitDiag}. Recalling that the azimuthal reflection symmetry is
represented by the sign change of the $\tilde{y}$ coordinate, we conclude that this
crisis bifurcation is a 
symmetry-restoring\cite{chossat1988symmetryincreasing} one at
which the previously-disconnected state-space regions that are reflection copies 
of one another merge to form the
final chaotic attractor (\figref{trajectories}{D}) of the pilot-wave system.
    The appearance of lemniscate-shaped trajectories (\figref{trajectories}D) 
    following the symmetry-restoring bifurcation suggests that it is the result 
    of a merger of oval-shaped chaotic trajectories with lemniscate ones, which lost 
    stability via boundary crisis at $\memory \approx 26.6$. This scenario, 
    illustrated as a state-space cartoon in \figref{cartoon}, is markedly different 
    from previously studied\cite{
        grebogi1982chaotic,grebogi1987critical,ott2002chaos} crisis bifurcations 
    that follow a chaotic attractor's collision with an unstable periodic orbit.

\begin{figure}
	\centering
	\includegraphics[width=\linewidth]{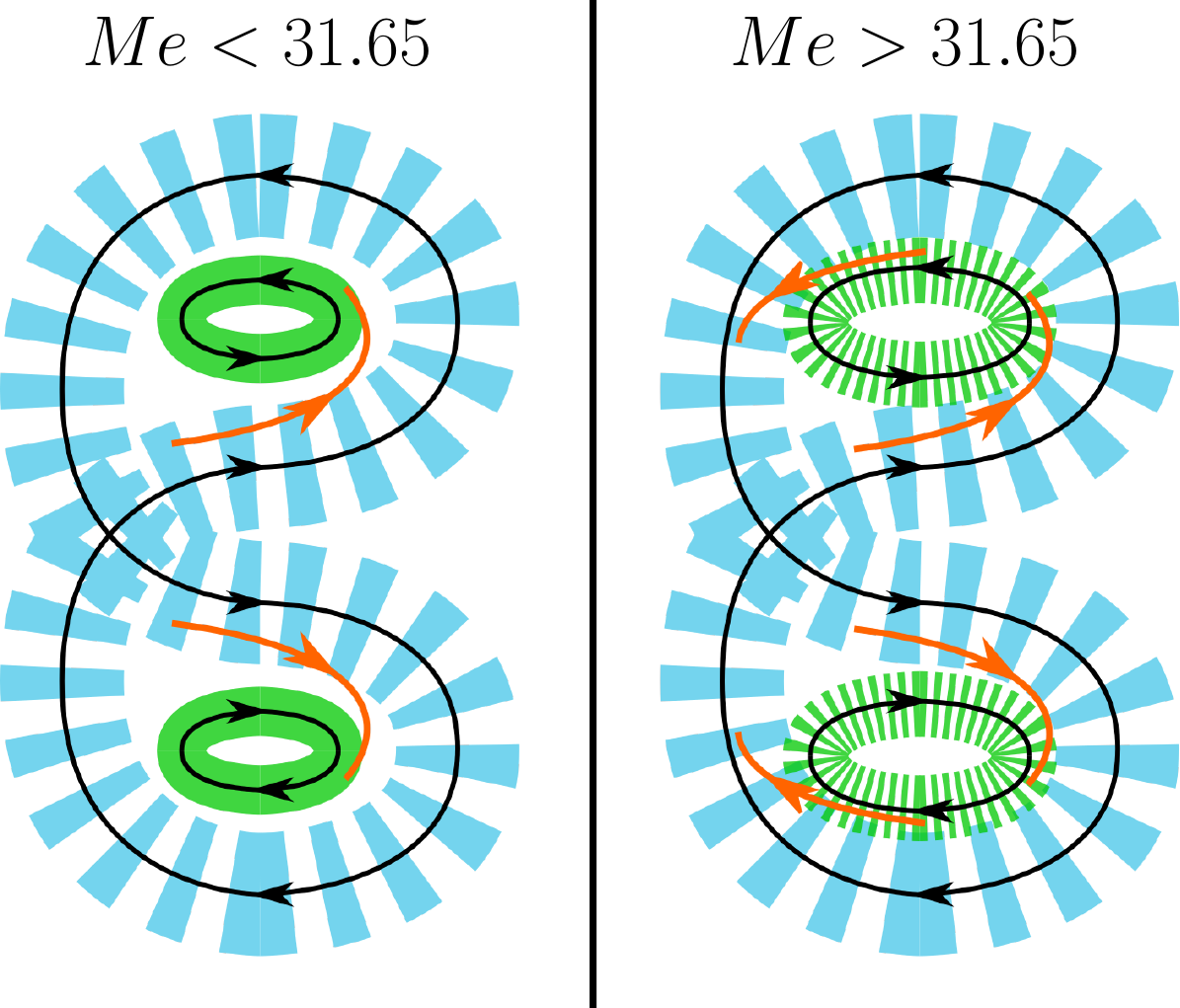}
	\caption{
            State space cartoon of the hydrodynamic pilot-wave system prior 
            ($Me < 31.65$) and after ($Me > 31.65$) the symmetry-increasing 
            crisis bifurcation. Solid (striped) areas correspond to distinct 
            stable (unstable) state space regions in the vicinity of lemniscate 
            and oval orbits depicted as closed curves. Orange curve segments 
            indicate possible transitions between the regions with arrowheads 
            showing the directions. 
	}
	\label{fig/cartoon}
\end{figure}

\subsection{Chaotic scattering}

In our bifurcation study, we observed seemingly chaotic
trajectories in the neighborhoods of both lemniscate and oval orbits. As
our final result and further evidence for the bifurcation scenario
depicted in \figref{cartoon}, we demonstrate that the system's dynamics
following the symmetry-increasing bifurcation at $\memory\approx31.65$ can be
understood as chaotic scatterings from the state-space regions 
corresponding to the lemniscate- and oval-shaped orbits. To begin, we
track the time-series of the angular momentum $L$ that
distinguishes clockwise and counter-clockwise rotating trajectories.
\figref{lifetimes}{A} shows the angular momentum time-series (blue)
corresponding to a chaotic trajectory segment similar to the one shown in
\figref{trajectories}{D}. Intervals, such as $\zeit \in [500 T_F, 1000 T_F]$,
during which the angular momentum oscillates between positive and negative
values of $L$ correspond to lemniscate-shaped orbits. Conversely, the episodes
during which the angular momentum remains either positive or negative correspond
to the oval-shaped trajectories. This indeed suggests that the lemniscate orbits
should separate the chaotic ovals with opposite senses of rotation, just prior
to the crisis bifurcation, and hence constitute a chaotic basin boundary.

If the chaotic attractor can be decomposed into distinct chaotic repellers,
then we would expect the droplet to spend exponentially-distributed times within each of
these sets.\cite{lai2011transient} In order to test this hypothesis, we first computed the
moving average $\langle L \rangle_{T_w}$ of the angular momentum,
shown in \figref{lifetimes}{A}, with a window length $T_w \in [104,113] T_F$ 
(see the appendix for details). During the lemniscate episodes, the 
window-averaged angular momentum remains near zero,
whereas for the ovals, it takes  a near constant value 
$\langle L \rangle_{T_w} \approx \pm 0.05 m \lambda_F^2 / T_F$. 
Following this observation, we choose the
thresholds $L_{l} = 0.025 m \lambda_F^2 / T_F$ and $L_{o} = 0.045 m \lambda_F^2
/ T_F$ and assume that the episodes with 
$|\langle L \rangle_{T_w}| < L_{l}$ (\figref{lifetimes}{A}, transparent red) and
$|\langle L \rangle_{T_w}| > L_{o}$ (\figref{lifetimes}{A}, transparent cyan) 
corresponds lemniscate and oval trajectories
respectively. Under this assumption, we estimated the distribution of the
lifetime $\tau$ for the lemniscate trajectories at four \memory\ values as shown
in \figref{lifetimes}{B}. The survival function $S(\tau)$ is the
probability of the droplet to remain in lemniscate-like motion for a time $\tau$,
and for all \memory\ values it appears to have an exponential tail for $\tau >
100 T_F$. Interestingly, this distribution shows very little variation upon
changing \memory , thus we fit an exponential to the mean of $S(\tau)$ in
\figref{lifetimes}{B} for $\tau \ge 125 T_F$. The lifetimes of the oval
trajectories shown in \figref{lifetimes}{C}, on the other hand, become
progressively shorter as we increase \memory . Similar to the lifetime
distributions of the lemniscates, the ovals also show exponential tails for high
$\tau$.

\begin{figure}
\begin{overpic}[width=\linewidth]{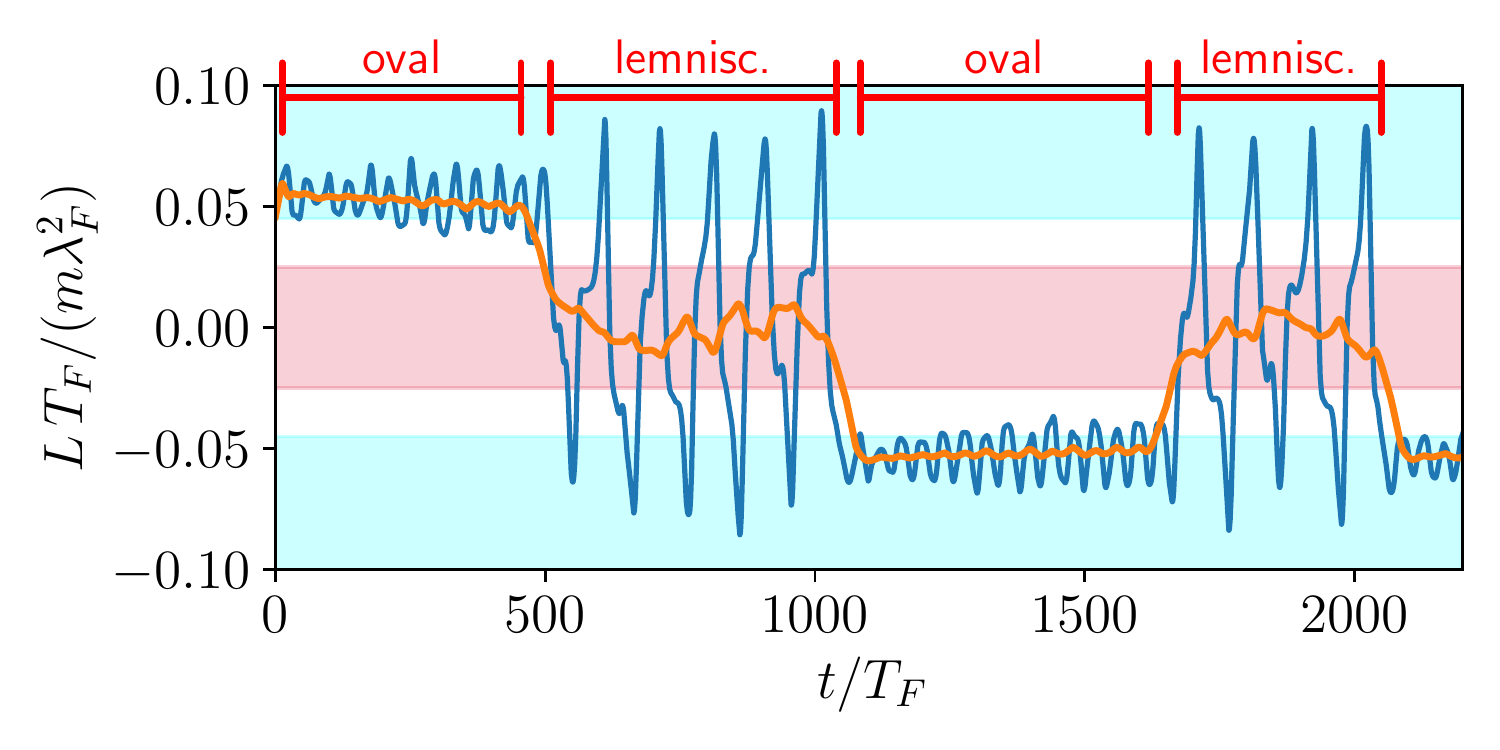}
	\put (2.1,45.0) {A.}
\end{overpic}	\\
\begin{overpic}[width=0.49\linewidth]{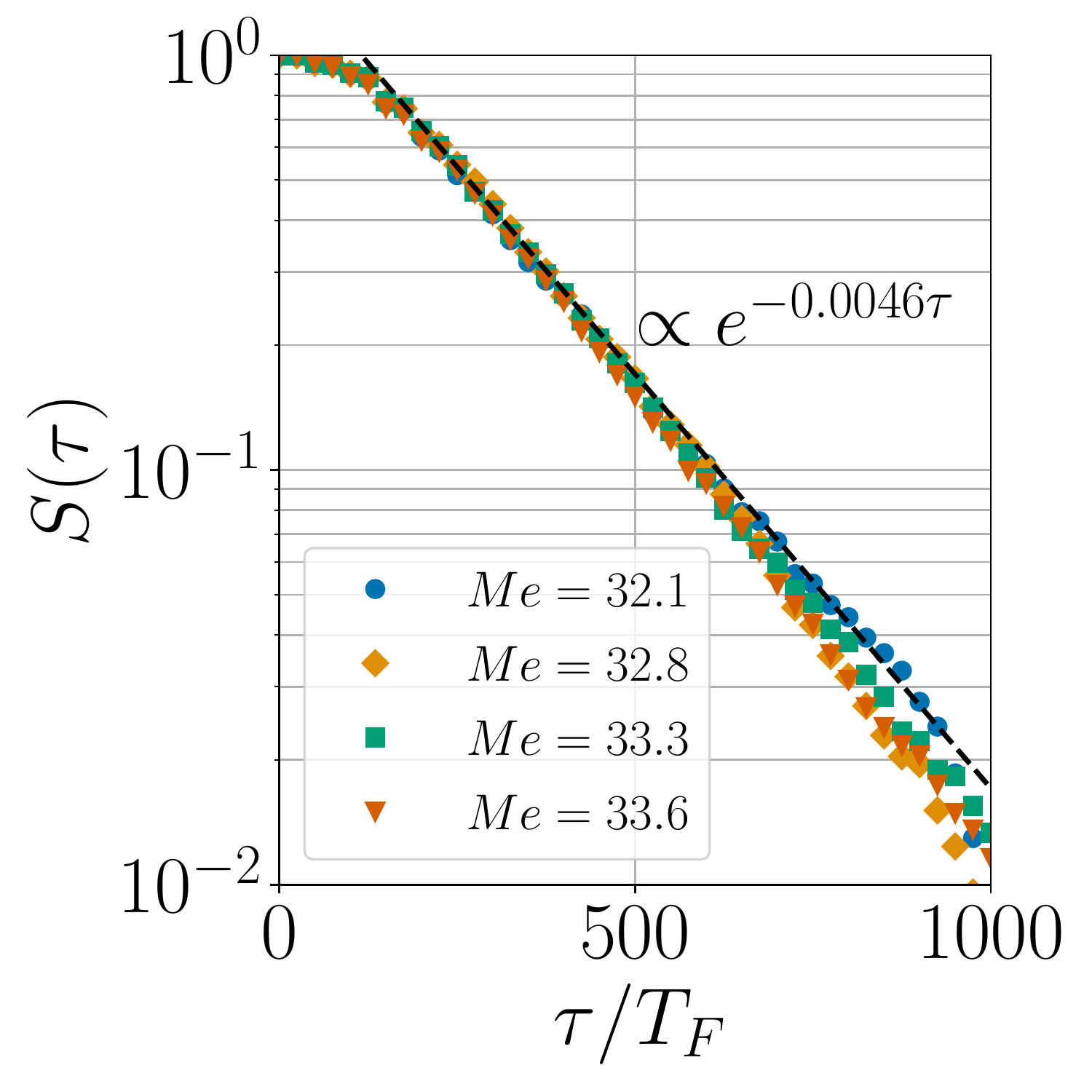}
	\put (3,95.0) {B.}
\end{overpic}
\begin{overpic}[width=0.49\linewidth]{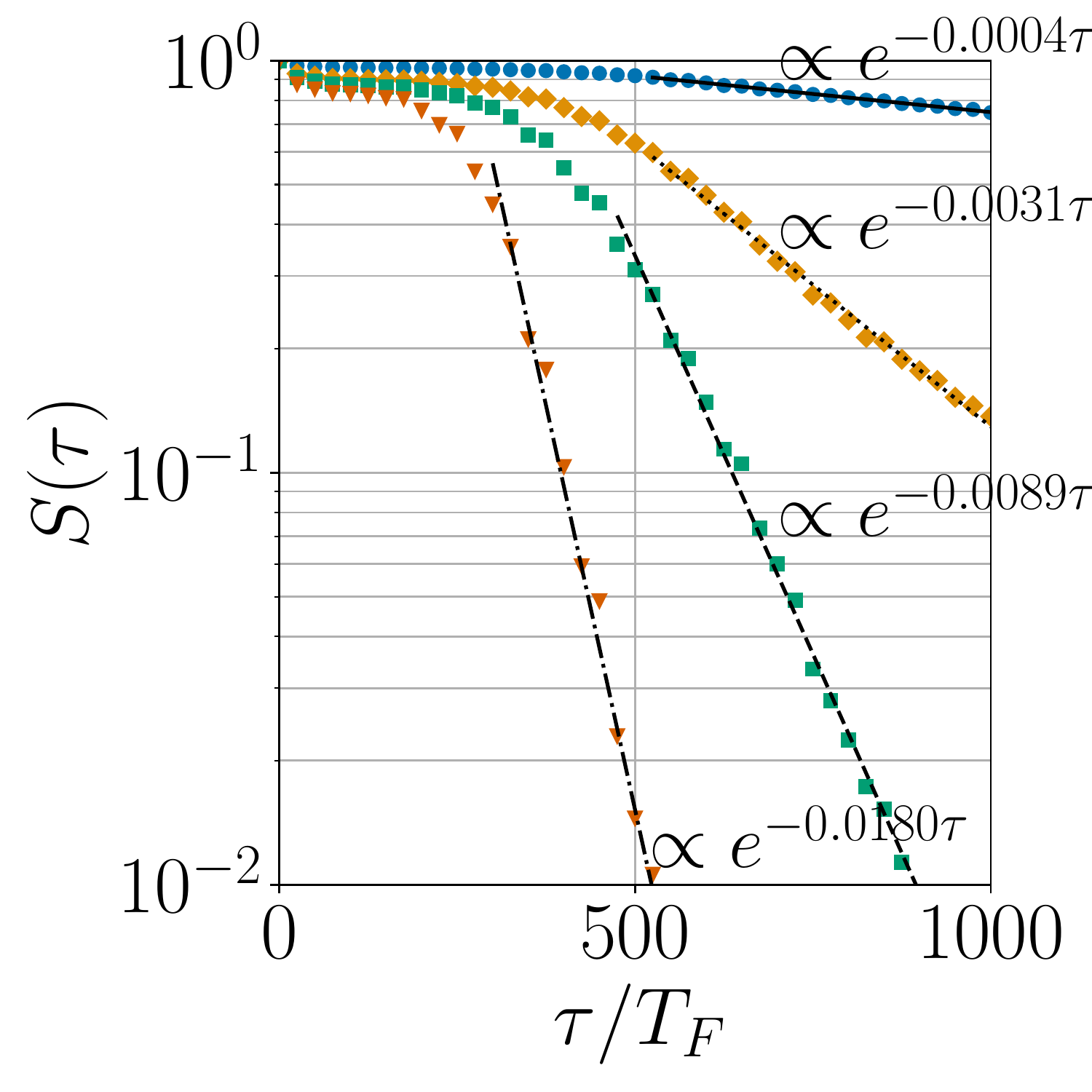}
	\put (3,95.0) {C.}
\end{overpic}
	\caption{
		(A) Time series of angular momentum (blue) and its window
		average (orange) for a chaotic droplet trajectory similar to
		the one shown in \figref{trajectories}{D}.
        The shaded red (cyan) indicates window-averaged angular momentum
        intervals that are ruled as lemniscate (oval) for computing the 
        lifetimes.
		(B) Lifetime distributions of the lemniscate trajectories
		at \memory\ values beyond the symmetry-increasing crisis
		along with an exponential fit to their average.
		(C) Lifetime distributions of the oval trajectories
		at \memory\ values beyond the symmetry-increasing crisis along
		with exponential fits (solid black) to the tails of the
		distributions.
		The legend for the \memory\ values of the data in
		C is identical to that in D, hence not shown.
}
\label{fig/lifetimes}
\end{figure}

\begin{figure}
	\begin{overpic}[width=0.49\linewidth]{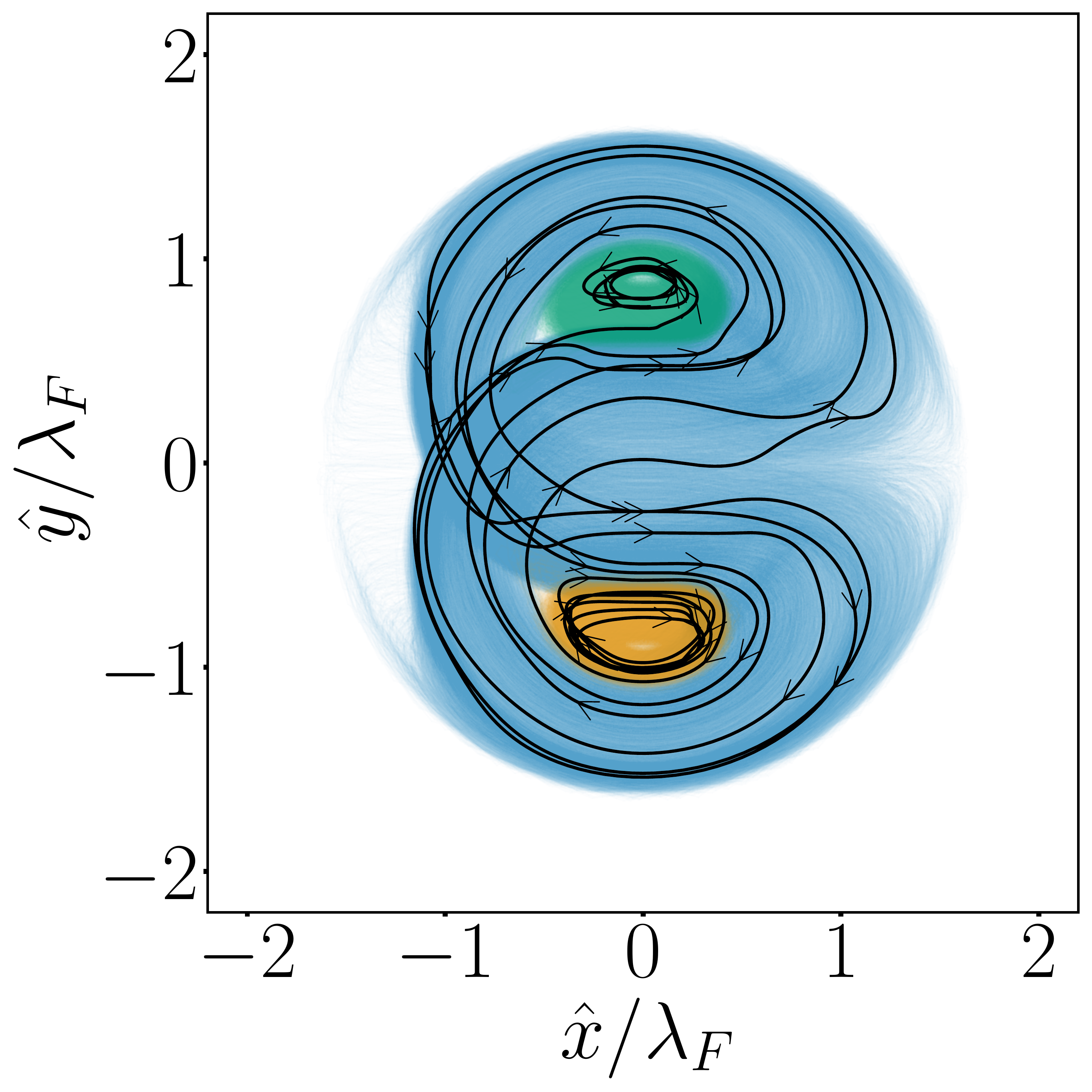}
		\put (3,95) {A.}
	\end{overpic}
	\begin{overpic}[width=0.49\linewidth]{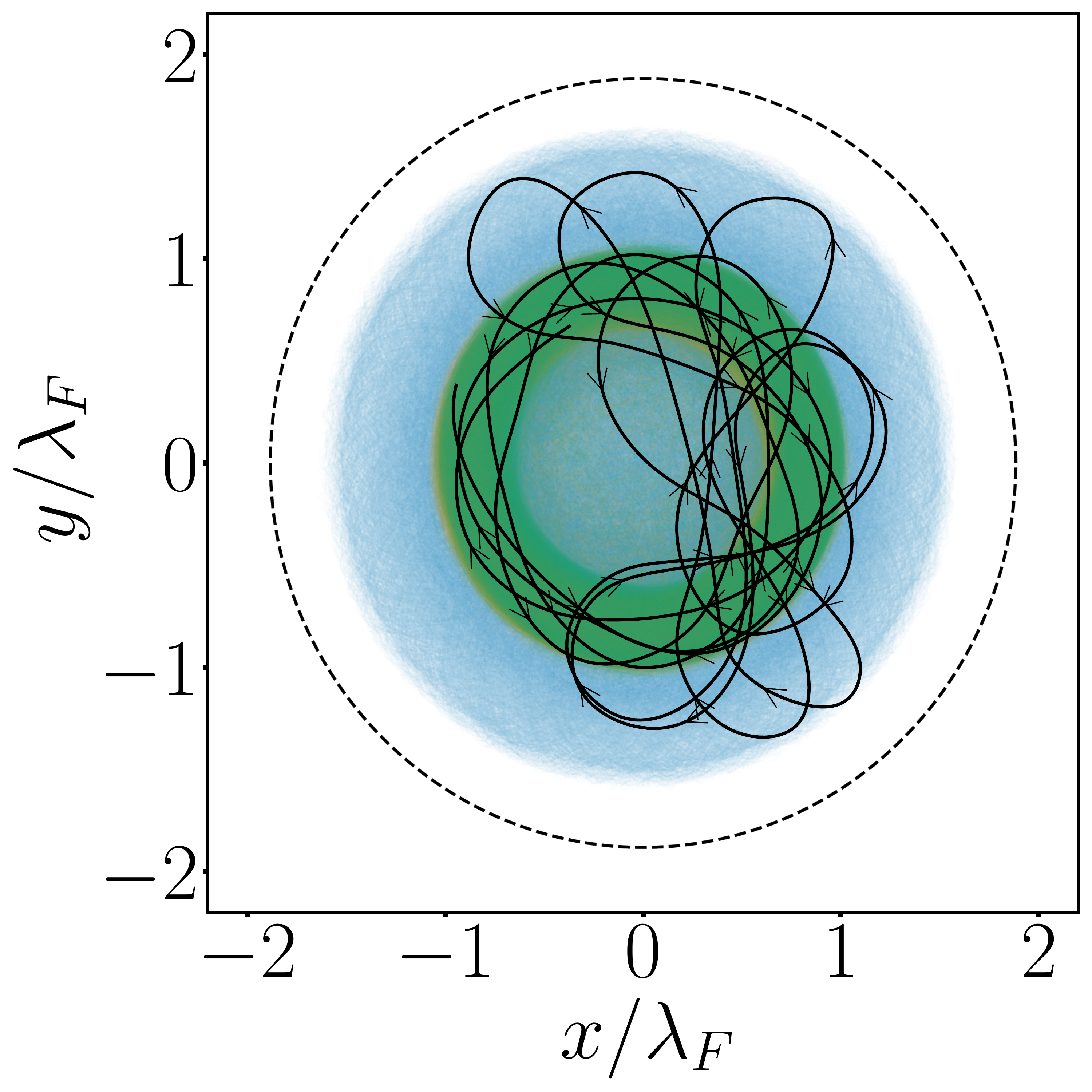}
		\put (3,95) {B.}
	\end{overpic}
	\caption{
		(A) Symmetry-reduced chaotic attractor of the pilot-wave
		system where colors correspond to points on the distinct chaotic
		repellers inferred from the window-averaged angular momentum
		of the trajectories.
		Blue points correspond to those on the lemniscate
		trajectories whereas the orange and green are those on
		ovals with $\langle L \rangle_{T_w} > L_{o}$
		and $\langle L \rangle_{T_w} < -L_{o}$ ,
		respectively.
		A trajectory (black) transitioning from the
		$\langle L \rangle_{T_w} < -L_{o}$ region (orange) to
		$\langle L \rangle_{T_w} > L_{o}$ region  (green)
		one after spending
		some time on the lemniscate repellor (blue)
		is also shown where arrowheads indicate the
		direction of time.
		(B) Same data points and the transitioning trajectory
		in the lab frame where the dashed circle indicates the
		corral boundary.
		Due to the degeneracy of lab-frame visualizations,
	  			 the orange and green data points overlap in B.
		\label{f-Transition}
		}
\end{figure}

A lifetime distribution with an exponential tail is the hallmark \cite{lai2011transient} 
of transient chaos.\cite{yorke1979metastable,pianigiani1979expanding,kadanoff1984escape}
Intuitively, one can understand the necessity of an 
exponential-tailed lifetime distribution by
arguing that a chaotic system is \textit{memoryless} \cite{papoulis1967probability} for
time scales much longer than the \textit{Lyapunov time} $\mu_L^{-1}$, where
$\mu_L$ is the leading Lyapunov exponent. Since any noise in the system is typically
amplified as $\exp{\mu_L \zeit}$, the system's memory of its present
state will be completely lost after a time $\zeit \gg \mu_L^{-1}$.
Therefore, the exponential tails of the lifetime distributions in
\figref{lifetimes}{B,C} lead us to the conclusion that the lemniscate- and
oval-shaped motions of the droplet correspond to distinct chaotic repellers
which are visited transiently by the dynamics. The overall
dynamics can, thus, be viewed as consecutive scatterings between these
strange repellers. In Fig.~\ref{f-Transition}A, we illustrate one such
scattering event where the droplet transitions from the oval-repeller with
positive angular momentum ($\langle L \rangle_{T_w} > L_{o}$) to the negative
side after spending some time on the lemniscate repeller. In order to 
illustrate these different regions in Fig.~\ref{f-Transition}, we also plotted samples 
from a very long trajectory with different colors corresponding to distinct 
state-space regions (blue: lemniscate, orange: oval with
$\langle L \rangle_{T_w} > L_{o}$, green: oval with $\langle L \rangle_{T_w} <
-L_{o}$). 

Prior to the symmetry-increasing crisis bifurcation at $\memory \approx 31.65$,
the oval-shaped chaotic trajectories constitute an attractor of our
pilot-wave system. The symmetry-increasing bifurcation is, therefore, one at
which the chaotic ovals lose stability and become transient. The post-crisis
reduction of the mean lifetimes of oval-shaped orbits (\figref{lifetimes}{C}) is
reminiscent of the well-understood behavior in dissipative two-dimensional
maps.\cite{grebogi1982chaotic,grebogi1987critical} In those systems, attractors
generically lose stability through homoclinic or heteroclinic tangencies of
periodic orbits. Subsequent mean lifetimes of the chaotic transients drop off
following a power-law as a function of the distance from the critical parameter
value at which the crisis bifurcation takes place. In the present case, our
measurements near the critical \memory\ are not dense enough to quantitatively
test a power-law behavior. Moreover, the chaotic attractor in the present case
loses stability through its merger with the chaotic lemniscates, for which, to
the best of our knowledge, there is no obvious reason to expect a power-law
scaling. Nevertheless, it is clear that the lifetimes of ovals become
progressively shorter as we increase the \memory\ further from its critical
value. We would like to note that this behavior can be exploited to tune the
probabilities of observing the droplets in oval and lemniscate states.

\section{Concluding remarks}\label{sec:remarks}

In this paper, we presented a parametric study of a hydrodynamic
pilot-wave system, in which we identified two crisis bifurcations. The first of
these occurs when chaotic lemniscate trajectories lose stability giving way to
stable circular-like trajectories. The second  symmetry-increasing one occurs
when a chaotic attractor corresponding to oval trajectories merges with the
lemniscate repeller, yielding the final attractor of the system. We have also
demonstrated that the proximity of the control parameter \memory\ to its
critical bifurcation value determines the lifetime distribution of the
oval-shaped motions, providing a useful tool for adjusting the global statistics
of similar pilot-wave systems. We note that global bifurcations due to the
merger of distinct chaotic sets were previously observed in the
discrete-\cite{rahman2016neimarksacker,rahman2017sigma,rahman2020interesting}
and continuous-time\cite{budanur2019state} models of the pilot-wave
hydrodynamics.

The idea of decomposing high-dimensional chaos into subunits of chaotic
repellers by identifying crisis bifurcations has recently been explored in the
numerical studies of transitionally turbulent shear
flows.\cite{kreilos2014increasing,zammert2015crisis,ritter2016emergence}
Similarly, chaotic basin boundaries were also investigated  in the numerical
studies of laminar-turbulence transition in shear
flows.\cite{schneider2007turbulence,schneider2010localized,zammert2014spotlike,budanur2018complexity}
The instability of these basin boundaries, however, renders them inaccessible to
direct observations in the laboratory. To the best of our knowledge, our study
is the first experimental demonstration of
transitions mediated by a chaotic basin boundary and the high-dimensional
attractor formation through the merger of distinct chaotic sets.
A future research direction that is motivated by these results is a theoretical 
treatment of high-dimensional attractors as those formed by chaotic 
repelleres around distinct periodic orbits. In particular, it might be possible 
to adapt the periodic orbit expansions for chaotic repellers \cite{kadanoff1984escape}
to predict lifetime distributions such as those shown in \figref{lifetimes}.

One of the most striking features of hydrodynamic pilot-wave systems is 
the wavelike patterns that are reminiscent of the quantum wave functions which 
emerge in long-time statistics of the droplet position at high 
\memory.\cite{harris2013wavelike,saenz2017statistical,saenz2020hydrodynamic}  
Already in Ref.\cite{harris2013wavelike}, Harris et al. argued that the 
emergent statistics can be understood as a droplet's transient visit of  
unstable (quasi-)periodic orbits. Our results are consistent with their 
insights and suggest new methods to understand the nature of deterministic 
dynamics underlying the emergent wavelike statistics. Specifically, 
lifetime measurements, such as those in \figref{lifetimes}, can be used to 
characterize individual neighborhoods in these experiments. Moreover, 
we would like
to note that tuning the lifetime distribution of a particular droplet motion by
varying a control parameter's proximity to a bifurcation value might be relevant
for quantum analogies since the adjustment of the lifetime distribution of some
state directly influences the probability of observing the droplet
in that state. In our case, we achieved this by varying \memory , which as we
moved farther away from the symmetry-increasing crisis bifurcation, resulted in
shorter and shorter oval lifetimes, hence a lower probability of observing it.

\begin{acknowledgments}
    This work was partially funded by the Institute of Science and
    Technology Austria Interdisciplinary Project Committee Grant ``Pilot-Wave
    Hydrodynamics: Chaos and Quantum Analogies''.
\end{acknowledgments}

\appendix*
\section{Experimental details}

The vibrating bath in the experiment is made from black anodized aluminum. The
circular corral located at its center has a diameter $19.95\pm 0.05\,\si{mm}$ and
depth  $6\pm 0.05\,\si{mm}$, as shown in \figref{setup}. The corral is
filled to a height $6.9\pm 0.1\,\si{mm}$ with silicone oil  (polydimethysiloxane),
which has kinematic viscosity $\nu ={21.5\times 10^{-6}}$\,m$^2$/s, density $\rho
= 953$\,kg/m$^3$, and surface tension  $\sigma$=20.8$\times 10^{-3}$\,\si{N/m} at
room temperature ($T$ = 21.6 $\pm 0.1$\,\si{\degreeCelsius}). For this fluid
layer depth, a thin $0.9\pm 0.1$\,\si{mm} overflow layer--that serves to dampen
surface waves--is formed outside the corral. A transparent plastic lid  placed
on top of the aluminum bath shields the corral (and the droplet) from stray air
currents.

The aluminum bath is mounted on an air-cooled electromagnetic shaker (Data
Physics V55) and is leveled perpendicular to gravity. The shaker oscillates
vertically at a frequency $f_0=75\,\si{Hz}$,  when a sinusoidally varying voltage
signal of the same frequency is input. Three piezoelectric sensors (PCB 352C65),
mounted as shown in \figref{setup}, measure the vertical ($\gamma$) and
horizontal ($\gamma_x,\gamma_y$) accelerations  of the vibrating aluminum bath.
The setup is aligned such that
$|\gamma_x|/|\gamma|,|\gamma_y|/|\gamma|\lesssim0.01$. For a given memory $Me$,
the target vertical acceleration $\gamma_{Me}$ is computed using
Eq.~\ref{e-Memory}, i.e., $\gamma_{Me} = \gamma_{F}(1-Me^{-1})$. Here, $\gamma_F
= 4.32\,g$ is the experimentally measured critical acceleration (in units of
$g=9.8\,$m$^2$/s) for the onset of  Faraday instability  at
$21.7\,\si{\degreeCelsius}$. A feedback loop controls the amplitude of the
sinusoidal voltage signal driving the shaker, such that the measured vertical
acceleration ($\gamma$) deviates from $\gamma_{Me}$ by less than $\pm0.1$\%,
i.e., $|\gamma-\gamma_{Me}|/\gamma_{Me} \leq 0.001$.

\begin{figure}[t!]
	\begin{overpic}[width=1.5in]{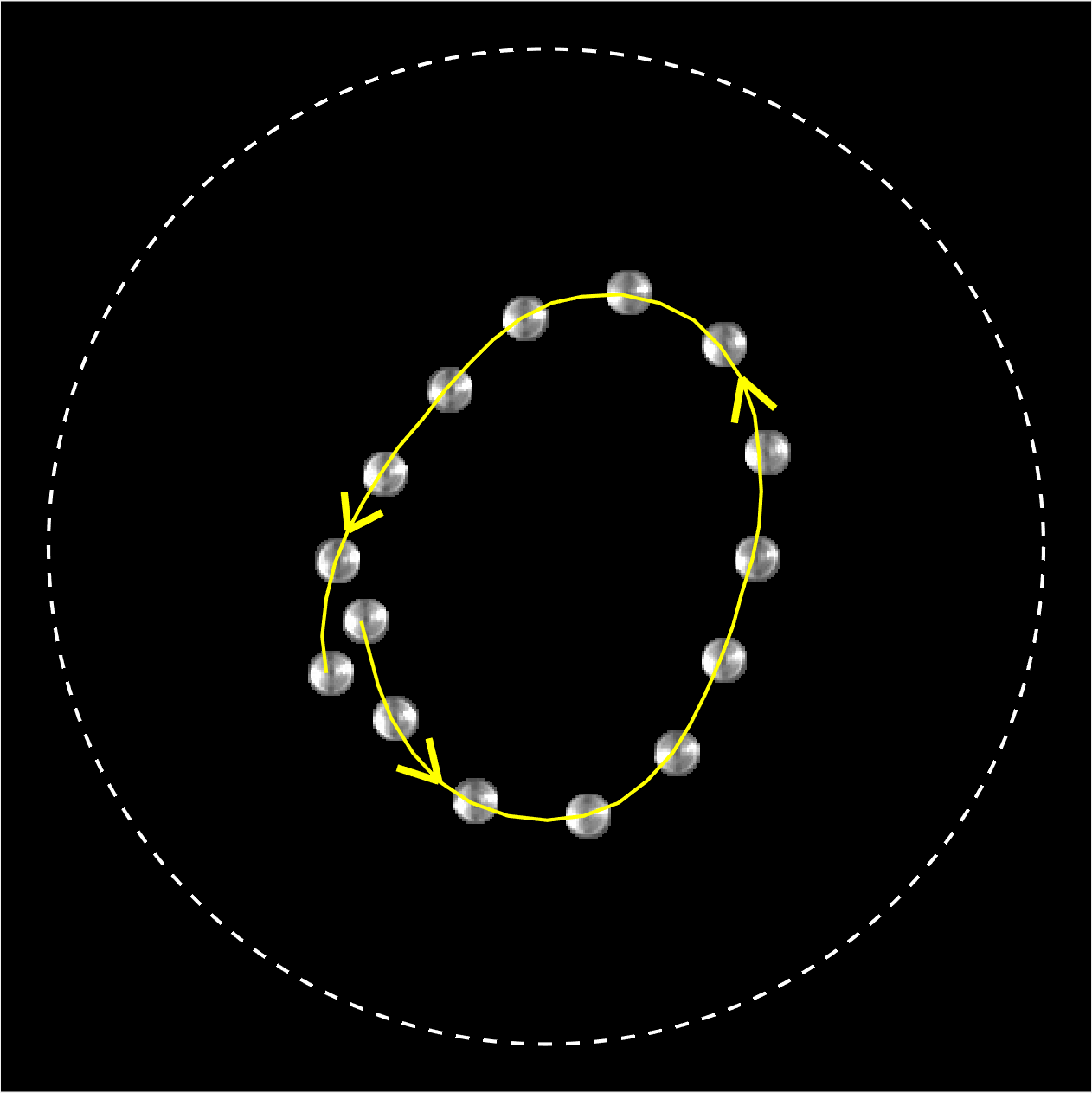}
		\put (1,93) {{\color{white} A.}}
	\end{overpic}
	\hspace{3mm}
	\begin{overpic}[width=1.5in]{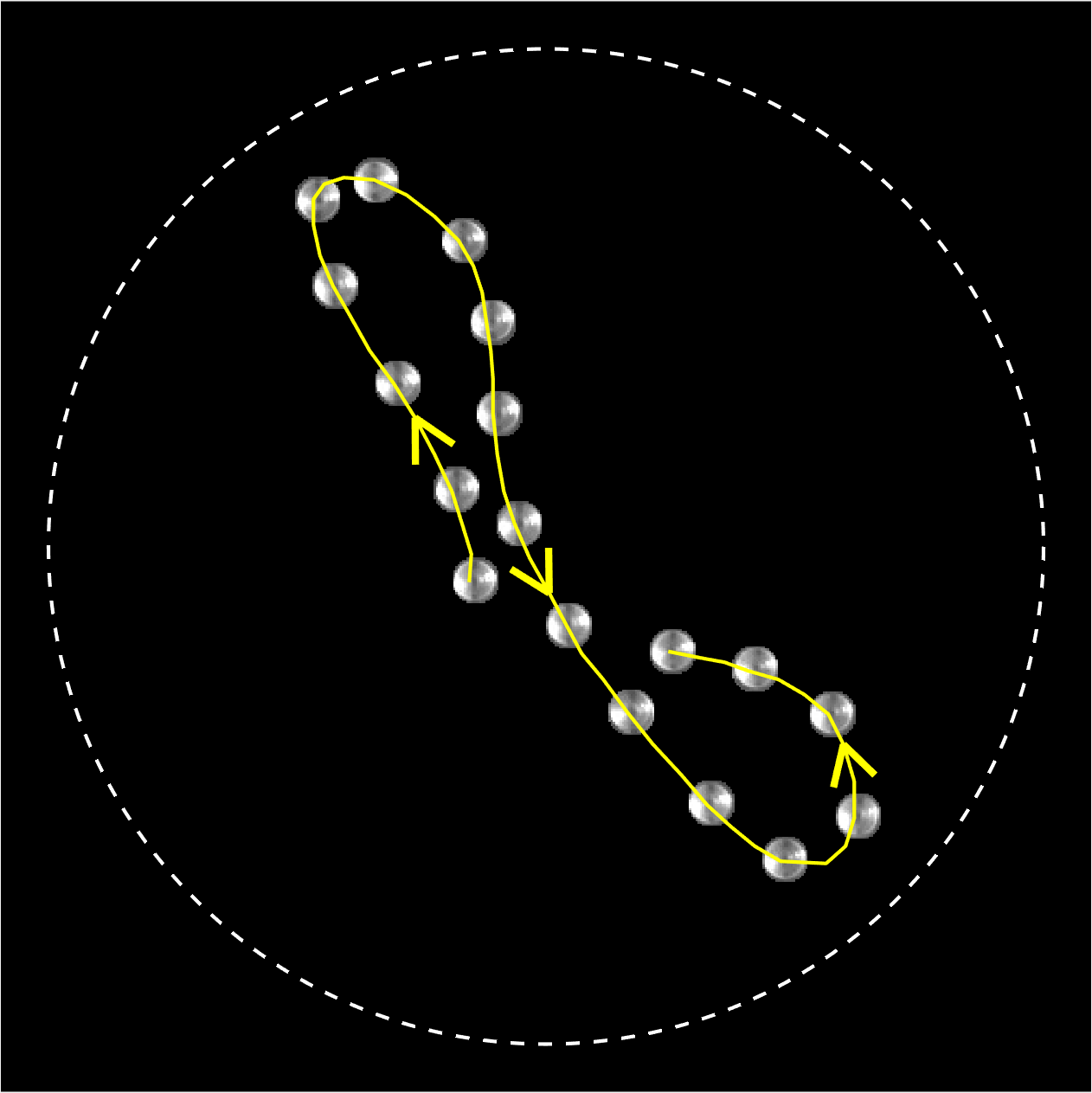}
		\put (1,93) {{\color{white} B.}}
	\end{overpic}	
	\caption{\label{fig:DropTraj} Reconstructed  trajectories of the silicone
    oil droplet tracing (A) Oval and (B) Lemniscate at $Me = 32.8$. The white
    dashed circle marks the boundary of the corral. The yellow curve is the
    trajectory of the brightest pixel (after Gaussian blurring) on the droplet.}
\end{figure}

To generate droplets of a desired size, silicone oil--filled in a syringe to a
fixed height--was drained through a 33 gauge needle for a fixed time. The
syringe was then touched against the vibrating bath to dislodge the droplet onto
the fluid surface. The longer (shorter) the duration of draining, that  larger
(smaller) is the drop size. For a fixed drain-time, droplets generated using
this technique varied in diameter by about $\pm 0.05\,$mm. All experimental runs
reported in this study were performed with a single silicone oil droplet of
diameter $0.85\pm0.05\,\si{mm}$. Images of the bouncing droplet were recorded at
intervals $\Delta t \approx 57$\,ms, using a CMOS camera (Basler acA2000-165um)
mounted above the bath (cf. \figref{setup}{B}). In pixel units, the diameters of
the corral and the droplet are 468$\pm$1 and 20$\pm$1, respectively. To track
the position of the droplet in real-time, a gaussian filter (with a 6 pixel
standard deviation) was applied to each image and the location of the brightest
pixel--approximating the coordinates  ($x_c,y_c$) of the droplet center--was
measured. The time series $x_c(t),y_c(t)$ were then interpolated onto a
temporal grid with spacing $\Delta t = 5.7$\,ms, using a cubic spline
interpolation. Instantaneous droplet velocities $v_x(t),v_y(t)$ were then
computed by computing derivatives of the spline interpolation. An an example,
Fig.~\ref{fig:DropTraj} shows droplet trajectories  reconstructed by overlaying
successive images, each  approximately $150$ milliseconds apart in time.

\begin{figure}
	\centering
	\includegraphics[width=3in]{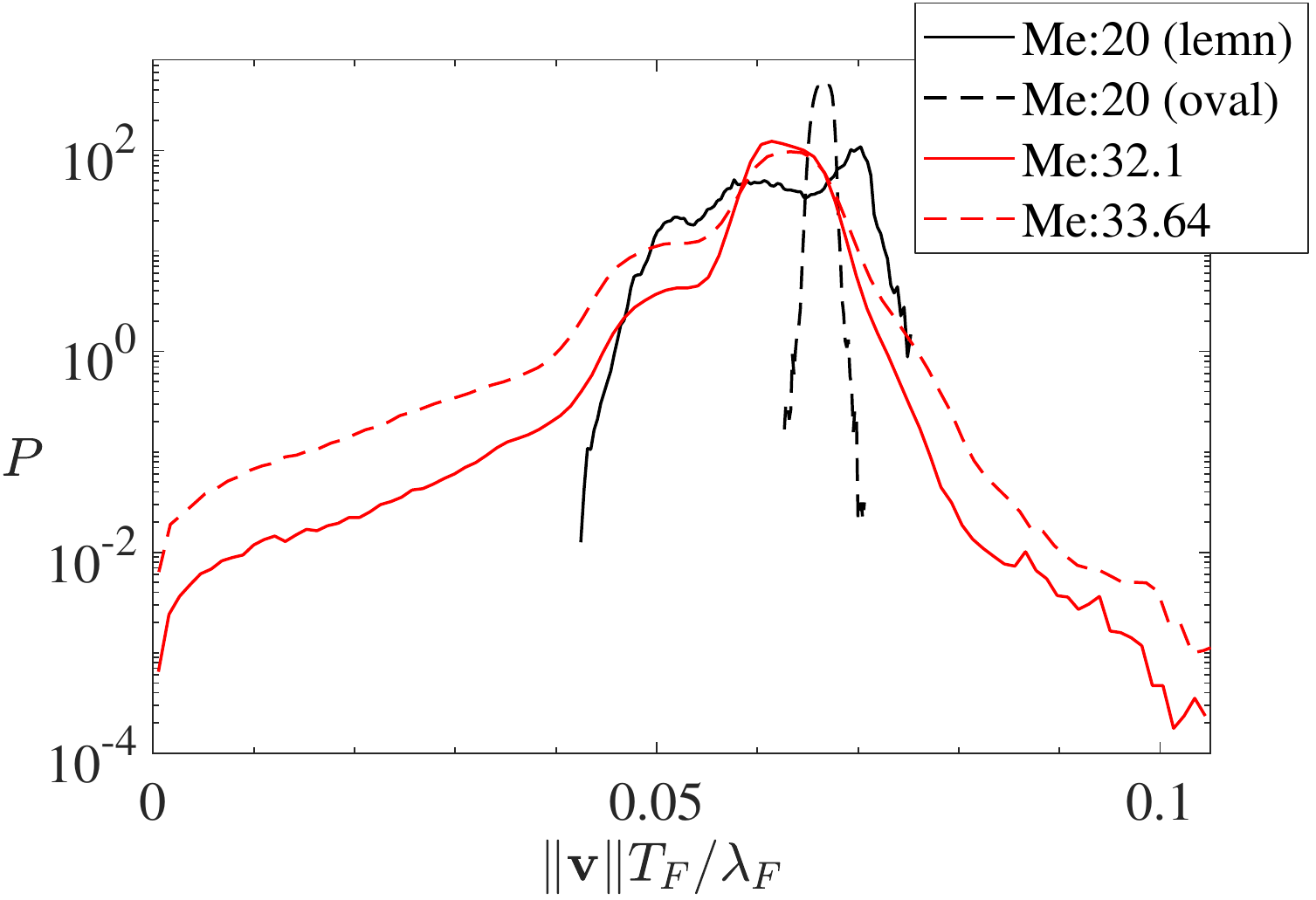}
	\caption{
        \label{fig:vpdf} Probability density function of the droplet velocity 
        at various $\memory$ values.
        }
\end{figure}

The horizontal speed $\|{\bf v}\|$ of the droplet remains fairly
constant ($\approx 0.06$), for the dynamical regimes explored in this
article. To demonstrate this, Fig.~\ref{fig:vpdf} shows that the probability
density function of $\|{\bf v}\|$ for representative values of $\memory \in
[20,33.64]$, normalized such that $\int P d\|{\bf v}\|=1$ in each case.
Recall that symmetry-reduction in the velocity plane requires that the speed
of the droplet does not vanish. Indeed, the probability of the droplet speed
$\|{\bf v}\|\leq 0.01$ is smaller than $10^{-3}$ for $\memory\geq 32.1$.
Even though this is an insignificant fraction, such decrease in droplet
speed is possibly due to the droplet (very rarely) approaching the boundary
of the inner corral and subsequently bouncing back. 
 
The sequence of bifurcations (Figs. \ref{fig/trajectories},
\ref{fig/scatterOrbitDiag}) was identified using an experimental run, where
memory $Me\in[20,33]$ was increased (decreased) in steps of $\Delta Me \approx
0.36$ ($\Delta Me \approx -0.37$) and the droplet position was tracked  for a
duration 1800\,s  at each $Me$. The lifetime distributions (cf.
\figref{lifetimes})  were estimated from separate runs, each approximately
$9\times 10^4$\,s long. The ambient room temperature (measured using a PT-100
probe placed beside the shaker) over the duration of these experimental runs was
fairly constant ($20.6\pm 0.1\,\si{\degreeCelsius}$). Nevertheless,  we measured
the critical memory $Me_c$ corresponding to the crisis bifurcation before and
after each experimental run.  $Me_c$ was reproducible to within $\Delta Me
\approx \pm 0.15$ for different runs, which suggests that the temperature of the
silicone oil, its physical properties, and consequently $\gamma_F$ do not vary
significantly across the various experiments runs. Lastly, although all
experimental runs reported in this article correspond to a single droplet, the
repeatability of results was validated using (at least five) different
experiments performed with different droplets and replacing the silicone oil in
the bath each time.

The survival probabilities in \figref{lifetimes}{B,C}, as mentioned in the
main text, were estimated from window-averaged angular momentum $\langle
L\rangle_{T_w}$ of the droplet. Here, $T_w$ is the average characteristic
time-scale for tracing the lemniscates and ovals at each value of \memory.
$T_w$ was estimated by computing the Fourier transform of the time-series of $L$
for the entire duration ($\approx 9\times 10^4$\,s) of each experimental run.
The following table lists the values of  $T_w$ for lemniscates and ovals at each
$\memory$.

\setlength{\tabcolsep}{7pt}
\begin{table}[!h]\centering
\caption{\label{table:Tw} {Characteristic times for tracing ovals and lemniscates.}}
\begin{tabular}{|c|c|c|c|c|}
\hline
{$Me$} & 32.1 & 32.8 & 33.3 & 33.6\\
{$T_w$ (lemniscate)} & 105 & 104 & 110 & 105\\
{$T_w$ (oval)} & 40& 39 &39 & 38\\
\hline
\end{tabular}
\end{table}

\bibliography{scatter.bib}{}
\bibliographystyle{unsrt}

\end{document}